\documentclass[showpacs,10pt,aps,prd,reprint,onecolumn,notitlepage,superscriptaddress]{revtex4-1}
\usepackage{amsmath}
\usepackage{bbm}
\usepackage{amsfonts}
\usepackage{amssymb}
\usepackage{bm}
\usepackage{tensor}
\usepackage{graphicx}
\usepackage[]{subfigure}
\usepackage{color}
\graphicspath{{/maybehome/rthompson/RTT2GB_local/Physics/Research/TransOptics/Figures/}}

\usepackage[]{hyperref}

\providecommand{\ed}{\mathrm{d}}
\providecommand{\mA}{\mathbf{A}}
\providecommand{\mF}{\mathbf{F}}
\providecommand{\mg}{\mathbf{g}}
\providecommand{\mG}{\mathbf{G}}

\providecommand{\mc}{\boldsymbol{\chi}}
\providecommand{\mlambda}{\boldsymbol{\Lambda}}

\begin{document}

\title{Cartographic distortions make dielectric spacetime analog models
imperfect mimickers}

\author{Mohsen Fathi}
\email{m.fathi@shargh.tpnu.ac.ir; \,\,mohsen.fathi@gmail.com}
\affiliation{Department of Physics, 
Payame Noor University (PNU),
P.O. Box 19395-3697 Tehran, Iran}

\author{Robert T. Thompson}
\email{rthompson@maths.otago.ac.nz \\ Authors appear in alphabetical order}
\affiliation{Department of Mathematics and Statistics,
University of Otago,
P.O.\ Box 56, 
Dunedin, 9054,  New Zealand
}

\begin{abstract}
It is commonly assumed that if the optical metric of a dielectric medium is identical to the metric of a vacuum space-time then light propagation through the dielectric mimics light propagation in the vacuum.
However, just as the curved surface of the Earth cannot be mapped into a flat plane without distortion of some surface features, so too is it impossible to project the behavior of light from the vacuum into a dielectric analog residing in Minkowski space-time without introducing distortions.
We study the covariance properties of dielectric analog space-times and the kinematics of a congruence of light in the analog, and show how certain features can be faithfully emulated in the analog depending on the choice of projection, but that not all features can be simultaneously emulated without distortion.  These findings indicate conceptual weaknesses in the idea of using analog space-times as a basis for transformation optics, and we show that a certain formulation of transformation optics closely related to analog space-times resolves these issues.

\bigskip

\noindent{\textit{keywords}}: artificial black hole, transformation optics, metamaterials

\end{abstract}

\maketitle

\section{Introduction}\label{sec:introduction}
Since ancient times, people have struggled with how best to draw a map.
The dawn of the Age of Discovery and its great seafaring voyages brought with it an urgent growth in cartographic investigation and prompted the development of differential and projective geometry.  
But it has been long recognized that all maps distort certain surface features, so that the question of which map to use is answered by which features one wants to represent and in which manner they should be represented.
Conformal maps such as the Mercator projection, prominent for use in navigation at sea, locally preserve angles and shapes at the expense of distorted areas - making Greenland look bigger than Africa and South America.
Other typical map projections preserve areas or distances between two points at the expense of distortions in other features.  
There are an infinite number of possible map projections, an infinite number that function as a compromise between distortions in two desired features, and an infinite number that do not preserve any particular feature.
Gauss' Theorema Egregium of 1828 illuminated the reason behind the inability to find a distortionless map, showing that any map between surfaces of different curvature will always introduce distortions.

In more recent times, there has been an interest in mimicking certain aspects of gravitating systems in laboratory settings.
For example, similarities between the space-time metric and the Navier-Stokes equations for a flowing fluid show that a sonic or surface wave horizon in the flowing fluid could provide an experimentally accessible surrogate for the study of near-horizon relativistic effects including Hawking radiation \cite{Hawking1975cmp,Unruh1981prl,Unruh1995prd,Weinfurtner2011prl}.
 
Light propagation through dielectric media is another system that could serve as such an analog space-time.
The connection between light propagation in gravitational systems and light propagation in dielectric media was  recognized in the early days of general relativity when it was observed that the relativistic deflection of starlight by a massive object such as the Sun could instead be explained by an appropriate distribution of refractive media around the object \cite{Eddington1920}.
It was subsequently shown by Gordon that the correspondence also holds in the other direction; that light propagation in certain refracting media can instead be mathematically described as light propagation through a curved vacuum space-time with an ``optical metric'' that is determined by the optical properties of the medium \cite{Gordon1923}.
Later, Plebanski identified the effective permittivity and permeability of an arbitrary space-time metric \cite{Plebanski1960pr}, and these results were used by de Felice to describe a dielectric representation of Schwarzschild space-time, arguably the first definite description of an analog black hole \cite{DeFelice1971gerg}.
In the last decade, developments in metamaterials \cite{Rotman1962iretrans,Schoenberg1983jasa,Pendry1996prl,Sievenpiper1996prl,Ziolkowski1997jap} have opened the possibility of physically realizing artificial media possessing some of the unusual dielectric properties required to construct a dielectric analog space-time and has sparked a resurgence of interest in the subject  \cite{Reznik2000prd,Schutzhold2002prl,Greenleaf2007prl,Narimanov2009apl,Thompson2010prd,Chen2010oe,Lu2010jap,Mackay2010pla,Miao2011plb,Smolyaninov2011jo,Smolyaninov2012prb,Bini2013gerg,Boston2015prd,FernandezNunez2016pla}.

Dielectric analog space-times are obtained as a projection of a curved vacuum manifold into a dielectric medium residing in a space-time of different curvature, usually flat Minkowski space-time, in very much the same manner that Earth's curved surface is cartographically projected onto a flat plane.
Gauss' Theorema Egregium applies equally to this modern concept of space-time map making; consequently, any dielectric analog space-time must necessarily introduce distortions into its representation of light propagation through the vacuum.
This fact seems to be underappreciated in much of the literature, which has largely focused on the form of the optical metric and the coordinate description of optical rays through the analog.
The majority of investigations in the subject have followed the footsteps of Plebanski and de Felice in identifying the dielectric properties of the analog and the optical metric with the vacuum metric, an identification that could mislead to the assumption that light propagation through the analog faithfully emulates the behavior of light propagating through the vacuum.

By analyzing the kinematics of a congruence of light in the analog, we show that the Plebanski analog fails to emulate physically measurable characteristics such as the fractional rate of expansion of a transverse cross section of the congruence.
Thus the Plebanski choice of projection map is actually quite similar to the Mercator projection in that it preserves the coordinate description of a ray but introduces distortions into the evolution of the cross-sectional area of a congruence.
As with cartography, there are an unlimited number of analog projections, and a more general mapping of vacuum space-times into analog representations was introduced in Ref.~\cite{Thompson2010prd}.
Using this, we demonstrate how an expansion-preserving projection map may be constructed, such that the expansion of a congruence is faithfully represented in the analog at the expense of distortions to the optical metric and the coordinate path of light rays.
We also provide a more detailed analysis of the covariance properties of dielectric analog space-times and show that the analog system lacks covariance in the sense that two analogs representing two different coordinate descriptions of the vacuum space-time are not themselves related by the same, or even an associated, coordinate transformation.

Furthermore, Plebanski's identification of a space-time metric with effective dielectric properties has been widely adopted as a basis for the interpretation of transformation optics (TO) \cite{Pendry2006sc,Leonhardt2006sc,Leonhardt2006njp1,Rahm2007pn}, and TO has been correspondingly invoked to undertake studies in analog space-times.
We argue that while dielectric analog space-times and TO are indeed very closely related, there are subtle differences that make the analog space-times approach unsuitable as a rigorous foundation for transformation optics.
Aside from conceptual issues, there are three problems with using analog space-times as the basis for TO.
First, the lack of any unique or canonical projection from the vacuum to the analog makes this approach to TO ill-defined.
Second, any useful formulation of TO should be fully covariant, while analog space-times suffer from covariance issues.
Third, the fact that no analog projection can faithfully replicate all aspects of light propagation in the vacuum means that the TO analog simply does not function as expected.

This paper is organized as follows. Section~\ref{sec:review} briefly reviews the covariant formulation of electrodynamics in dielectric media. Section~\ref{sec:dielctricAnalogs-review} reviews a rigorous theory of dielectric analog space-times and studies their covariance properties. In Sec.~\ref{sec:LightPropagation} we study the kinematics of light propagation through the analog, with a particular focus on the transverse expansion, and show how an expansion-preserving analog projection map may be constructed for a particular choice of congruence.
In Sec.~\ref{sec:Misconceptions} we compare analog space-times with TO, arguing that while they are closely related they are sufficiently, though subtly, distinct to have important consequences. Concluding remarks are given in Sec.~\ref{sec:conclusion}.  Throughout the paper we provide examples and illustrative calculations in terms of the Friedmann-Lama\^{i}tre-Robertson-Walker (FLRW) analog space-time.  The metric signature is chosen to be $(-+++)$ and the speed of light in vacuum is set to $c=1$.

\section{Classical Electrodynamics in Covariant Form}\label{sec:review}
We wish to describe electrodynamics inside dielectric media residing in a possibly curved space-time $M$ with metric $\mg$.
We use a tensorial, exterior calculus formulation of electrodynamics in dielectric media, the principal notions of which are reviewed here.  
A more complete introduction to tensorial electrodynamics may be found in, e.g., Refs.~\cite{Post,MTW,Baez}, while a more complete discussion of our application to electrodynamics in media may be found in Ref.~\cite{Thompson2012aiep}.

In this formulation, components of the vector field pairs $(\vec{E},\vec{B})$ and $(\vec{D},\vec{H})$ are seen to be components of the field strength and excitation 2-forms (antisymmetric second-rank covariant tensors) $\mF = F_{\mu\nu}$ and $\mG = G_{\mu\nu}$.
In a locally flat Cartesian frame, $\mF$ and $\mG$ have the matrix representations
\begin{equation}\label{Eq:ComponF}
F_{\mu\nu}=\begin{pmatrix}
  0 & -E_x & -E_y & -E_z \\
  E_x & 0 & B_z & -B_y \\
  E_y & -B_z & 0 & B_x \\
  E_z & B_y & -B_x & 0 
\end{pmatrix},
\end{equation}
and
\begin{equation}\label{Eq:ComponG}
G_{\mu\nu}=\begin{pmatrix}
  0 & H_x & H_y & H_z \\
  -H_x & 0 & D_z & -D_y \\
  -H_y & -D_z & 0 & D_x \\
 -H_z & D_y & -D_x & 0 
\end{pmatrix}.
\end{equation}

Pseudo-Riemannian $m$-dimensional manifolds such as $M$ are naturally endowed with an operator called the Hodge dual that for $k<m$ takes a $k$-form to an $(m-k)$-form.
For the purposes pursued herein, we only require the Hodge dual to act on 2-forms.
In the case of a four-dimensional space-time, the Hodge dual maps a 2-form into another 2-form and has the following index expression:
\begin{equation}\label{Eq:HodgeCompon}
\star\indices{_{\alpha\beta}^{\mu\nu}}=\frac{1}{2}\sqrt{|g|}\,\,\epsilon_{\alpha\beta\sigma\rho}g^{\sigma\mu}g^{\rho\nu}.
\end{equation}
Hence, the components of $\star\mathbf{F}$ in a local Cartesian frame are
\begin{equation}\label{Eq:StarFCompon}
(\star\mathbf{F})_{\alpha\beta}=\frac{1}{2}\sqrt{|g|}\,\,\epsilon_{\alpha\beta\sigma\rho}g^{\sigma\mu}g^{\rho\nu}F_{\mu\nu}=\begin{pmatrix}
  0 & B_x & B_y & B_z \\
  -B_x & 0 & E_z & -E_y \\
  -B_y & -E_z & 0 & E_x \\
  -B_z & E_y & -E_x & 0 
\end{pmatrix}.
\end{equation}

Maxwell's homogeneous and inhomogeneous equations are
\begin{subequations}\label{Eq:Maxwell} 
\begin{align}
\ed\mF&=0,\label{Eq:Hom}
\\
\ed\mG&=\boldsymbol{\mathcal{I}},\label{Eq:InHom}
\end{align}
\end{subequations}
where $\ed$ denotes the exterior derivative and $\boldsymbol{\mathcal{I}}$ is the charge-current 3-form, though in what follows we assume $\boldsymbol{\mathcal{I}}=0$. 
A complete solution of Maxwell's equations requires the specification of a constitutive relation between the fields $\mF$ and $\mG$.
It has been shown that such a constitutive relation for linear media has the form \cite{Thompson2011jo1,Thompson2011jo2}
\begin{equation}\label{Eq:G=StarF}
  \mathbf{G}=\star\mc\mF,
\end{equation}
where $\mc$ contains all the information about the medium such as permeability and permittivity, while $\star$ separately contains information about the space-time geometry.
In component form, the constitutive relation reads
\begin{equation}\label{Eq:G=StarFCompon}
   G_{\mu\nu}=\star\indices{_{\mu\nu}^{\alpha\beta}}\chi\indices{_{\alpha\beta}^{\sigma\rho}}F_{\sigma\rho}.
\end{equation}
The dielectric tensor $\chi\indices{_{\mu\nu}^{\alpha\beta}}$ is independently
antisymmetric under index exchange on either $\mu\nu$ or
$\alpha\beta$, and 
thus $\mc$ has a maximum of 36 independent parameters.  In practice, further symmetries may be imposed based on thermodynamic or energy conservation arguments that reduce the number of independent components to 15. 
The coordinate-independent vacuum $\mc_{vac}$ is uniquely defined as the trivial dielectric such that $\mc_{vac}\mF=\mF$.

Component-wise expansion of the constitutive Eq.~(\ref{Eq:G=StarFCompon}) in a local Cartesian frame recovers the vectorial relations \cite{Thompson2011jo2}
\begin{equation}\label{Eq:Constitutive1}
   \vec{D}=\bar{\bar\varepsilon}^c \vec{E} + \tensor[^b]{\bar{\bar{\gamma}}}{^c} \vec{B},  \quad\vec{H}=\bar{\bar\mu}^c\vec{B} + \tensor[^e]{\bar{\bar{\gamma}}}{^c}\vec{E}
\end{equation}
which can be put into the traditional relations
\begin{equation}\label{Eq:Constitutive2}
    \vec{D}=\bar{\bar\varepsilon}\vec{E} + \tensor[^h]{\bar{\bar{\gamma}}}{}\vec{H}, \quad\vec{B}=\bar{\bar\mu}\vec{H}+ \tensor[^e]{\bar{\bar{\gamma}}}{}\,\vec{E}
\end{equation}
where the ``traditional''  $3\times3$ matrices of permittivity $\bar{\bar\varepsilon}$,
permeability $\bar{\bar\mu}$, and magnetoelectric couplings
$^h\bar{\bar\gamma}$ and $^e\bar{\bar\gamma}$ in Eq.~(\ref{Eq:Constitutive2}), are
related to the ``covariant'' quantities of Eq.~(\ref{Eq:Constitutive1}) as follows:
\begin{subequations}\label{Eq:Material} 
\begin{align}
\bar{\bar\mu}&=\left(\bar{\bar\mu}^c\right)^{-1}, \quad \bar{\bar\varepsilon}=\bar{\bar\varepsilon}^c-\left(\tensor[^b]{\bar{\bar{\gamma}}}{^c}\right)\left(\bar{\bar\mu}\right)\,\left(\tensor[^e]{\bar{\bar{\gamma}}}{^c}\right)
,\label{Eq:mu-epsilon}
\\
\tensor[^h]{\bar{\bar{\gamma}}}{} &= \left(\tensor[^b]{\bar{\bar{\gamma}}}{^c}\right)\,\left(\bar{\bar{\mu}}\right),\quad \tensor[^e]{\bar{\bar{\gamma}}}{} =-\left(\bar{\bar{\mu}}\right)\,\left(\tensor[^e]{\bar{\bar{\gamma}}}{}\right). \label{Eq:gamma}
\end{align}
\end{subequations}
Thus in a locally flat frame, the components of $\mc$ may be identified as the usual permeability and permittivity, but the meaning of these concepts is less clear in the general curvilinear coordinates of curved space-times where the metric contributes nontrivially to the constitutive relation.

\section{Dielectric Analog of a space-time}\label{sec:dielctricAnalogs-review}
Equations (\ref{Eq:Maxwell}) and (\ref{Eq:G=StarF}) provide a description of electrodynamics in linear, dispersionless, lossless media residing in possibly curved space-time manifolds.
By selecting a map from one such manifold to another, the behavior of electromagnetic fields in one manifold may be identified with electromagnetic fields in another manifold such that Maxwell's equations are satisfied in the target manifold.
In particular, on-shell electromagnetic fields in an arbitrary vacuum space-time may be identified with on-shell solutions in a nonvacuum dielectric residing in Minkowski space-time, thereby enabling a flat space-time dielectric representation of a curved space-time in much the same way that the curved surface of the Earth may be mapped into a flat cartographic representation.

But as the map-makers of old discovered, there is no canonical choice of projection map.  
Indeed, since the curved space-time and the flat space-time are not isometric, Gauss' Theorema Egregium demands that not all features of the curved space-time can be simultaneously faithfully represented in the flat space-time, and the choice of projection map is informed by what features of the curved space-time one wishes to faithfully represent.
Choosing a coordinate identity as the projection map recovers Plebanski's early results \cite{Plebanski1960pr} that have been used almost exclusively in the study of dielectric analogs ever since.

\begin{figure}[htp]
 \includegraphics[]{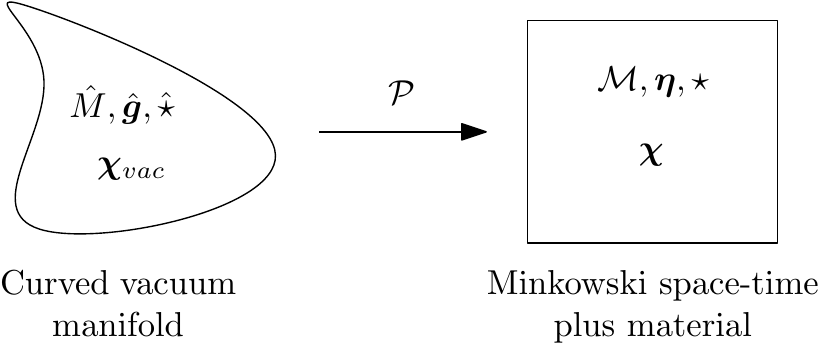}
  \caption{An arbitrary curved vacuum manifold $\hat{M}$ is projected into a dielectric analog $\mc$ residing in Minkowski space-time $\mathcal{M}$.  Electromagnetic fields on $\hat{M}$ are projected to $\mc$ by the pullback of $\mathcal{P}^{-1}$.}
  \label{Fig:AnalogMap}
\end{figure}

Consider the curved vacuum space-time manifold $\hat{M}$ with metric $\hat{\mg}$.  
The dielectric properties of this vacuum space-time are uniquely described by $\hat{\mc}=\mc_{vac}$.  
We wish to project on-shell solutions of Maxwell's equations in $\hat{M}$ into on-shell solutions of Maxwell's equations in a dielectric medium $\mc$ residing in Minkowski space-time $\mathcal{M}$ with metric $\boldsymbol{\eta}$, as in Fig.~\ref{Fig:AnalogMap}.
For this purpose, define the \textit{analog projection map}
\begin{equation}
 \mathcal{P}: \hat{M} \to \tilde{M}\subseteq\mathcal{M}.
\end{equation}

The electromagnetic fields $\mF$ and $\mG$ are projected into the analog with the pullback of $\mathcal{P}^{-1}$, which is denoted $(\mathcal{P}^{-1})^*$.
Mapping on-shell solutions from $\hat{M}$ to $\mc$ requires satisfaction of the two conditions
\begin{subequations} \label{Eq:G=pll}
\begin{align}
  \mG_x=\left(\mathcal{P}^{-1}\right)^*\left(\hat{\mG}_{\mathcal{P}^{-1}(x)}\right)=\left(\mathcal{P}^{-1}\right)^*\left(\hat{\star}_{\mathcal{P}^{-1}(x)}\circ\hat{\mc}_{\mathcal{P}^{-1}(x)}\circ\hat{\mF}_{\mathcal{P}^{-1}(x)}\right),\label{Eq:G=plla}
\\
 \mG_x = \star_x\circ\mc_x\circ\mF_x = \star_x\circ\mc_x\circ \left(\mathcal{P}^{-1}\right)^*\left(\hat{\mF}_{\mathcal{P}^{-1}(x)}\right)\label{Eq:G=pllb}.
\end{align}
\end{subequations}
Equations~(\ref{Eq:G=plla}) and  (\ref{Eq:G=pllb}) can be solved for the analog medium (setting $\hat{\mc}=\mc_{\mathrm{vac}}$), resulting in \cite{Thompson2010prd}
\begin{equation}\label{Eq:chi-general}
  {\chi}\indices{_{\eta\tau}^{\pi\theta}}(x)
       =-\star\indices{_{\eta\tau}^{\lambda\kappa}}|_x~\left(\Lambda^{-1}\right)\indices{^{A}_{\lambda}}|_x~\left(\Lambda^{-1}\right)\indices{^{B}_{\kappa}}|_x~\hat{\star}\indices{_{AB}^{MN}}|_{\mathcal{P}^{-1}(x)}~\Lambda\indices{^{\pi}_{M}}|_x~\Lambda\indices{^{\theta}_{N}}|_x,
\end{equation}
where $\mlambda = \ed\mathcal{P}$ is equivalent to the Jacobian matrix of the projection $\mathcal{P}$.
Capital latin indices refer to the vacuum manifold, while lower case greek indices refer to the Minkowskian analog.
Equation~(\ref{Eq:chi-general}) gives the parameters of a dielectric material $\mc$ in Minkowski space-time that supports the electromagnetic field configuration specified by the projection $\mathcal{P}$. 
In this sense, the medium $\mc$ emulates the behavior of electromagnetic fields in the curved vacuum space-time.

Choosing $\mathcal{P} = \mathcal{P}_{id}=id$, the coordinate identity map, then $\mlambda = \mathbb{I}_4$ and
\begin{equation}\label{Eq:TrivialAnalog}
  {\chi}\indices{_{\eta\tau}^{\pi\theta}}(x)
       =-\star\indices{_{\eta\tau}^{\lambda\kappa}}|_x~\hat{\star}\indices{_{\lambda\kappa}^{\pi\theta}}|_x.
\end{equation}
Expanding the constitutive Eq.~(\ref{Eq:G=StarF}) for this trivial analog medium in Minkowski space-time and converting to the representation of Eq.~(\ref{Eq:Constitutive2}) leads directly to Plebanski's results for the effective permeability, permittivity, and magnetoelectric couplings of a vacuum space-time \cite{Plebanski1960pr}
\begin{equation}\label{Eq:Plebanskiepsilon}
 \bar{\bar{\mu}}=\bar{\bar{\varepsilon}}=-\frac{\sqrt{-g}}{g_{00}}~g^{ij},
\end{equation}
\begin{equation}\label{Eq:magnetoelectric}
\tensor[^e]{\bar{\bar{\gamma}}}{}=\left(\tensor[^h]{\bar{\bar{\gamma}}}{}\right)^T=-\epsilon_{ijk}\frac{g_{0j}}{g_{00}},
\end{equation}
where $g^{ij}$ is the purely spatial part of the vacuum space-time metric and $g_{0j}$ are the time-space components of the metric.
If the permeability and permittivity are isotropic, then the bianisotropy vector evident in the magnetoelectric coupling may be identified with a characteristic velocity of the medium
\begin{equation}\label{Eq:bianisotropy}
v_j=\frac{g_{0j}}{g_{00}}.
\end{equation}
Note that Eqs.~(\ref{Eq:Plebanskiepsilon})-(\ref{Eq:bianisotropy}) do not conserve index type since they are not a covariant representation of the medium.

\subsection{Covariance properties of analog space-times}
Analog space-times are not covariant in the sense that they depend on both the choice of coordinates in $\hat{M}$ and the choice of projection, and a coordinate transformation in $\hat{M}$ does not correspond to a coordinate transformation in the analog.
 
On the one hand it is straightforward to see that Plebanski's analog model, based on the projection $\mathcal{P}_{id}$, leads to analogs that depend on the choice of coordinates in $\hat{M}$.
Given two coordinate descriptions of $\hat{M}$ and employing $\mathcal{P}_{id}$ on each, two analog mediums are calculated and found to be physically different objects rather than different coordinate descriptions of the same physical object. 
For example, consider FLRW space-time in comoving coordinates $(t,r_c,\theta,\phi)$ and physical coordinates $(t,r_p,\theta,\phi)$, related by $r_p= a(t)r_c$.
If both the comoving and physical coordinate descriptions of FLRW are projected into the spherical coordinates of Minkowski space-time with $\mathcal{P}_{id}$, then $\mc_c \neq \mc_p$ are both already described in the same Minkowski spherical coordinates and cannot be related by a coordinate transformation.
In particular, suppose $J\indices{^{A'}_{A}}$ is the Jacobian matrix of a coordinate transformation $\mathcal{R}:\hat{M}\to\hat{M}'$ in the vacuum space-time.  Then Eq.~(\ref{Eq:TrivialAnalog}) would become
\begin{equation} \label{Eq:PlebanskiNonCovariance}
   {\chi}\indices{_{\eta\tau}^{\pi\theta}}(x)
        =-\star\indices{_{\eta\tau}^{\lambda\kappa}}|_x~J\indices{^{A'}_{\lambda}}|_x~J\indices{^{B'}_{\kappa}}|_x~ \hat{\star}\indices{_{A'B'}^{C'D'}}|_{\mathcal{R}(x)}~ (J^{-1})\indices{^{\pi}_{C'}}|_x~(J^{-1})\indices{^{\theta}_{D'}}|_x,
\end{equation}
where the unprimed index on $\bm{J}$ is equivalent to an index in the analog because of the fixed projection map $\mathcal{P}_{id}$.  
If the first two factors of $\bm{J}$ could be pulled through the leading $\star$ operator such that the $\star$ acted on $\hat{\star}$ as in Eq.~(\ref{Eq:TrivialAnalog}), then Eq.~(\ref{Eq:PlebanskiNonCovariance}) would show that $\mc$ is related to $\mc'$ by a coordinate transformation.  
But this is clearly not possible so we conclude that analog space-times based solely on $\mathcal{P}_{id}$ are not covariant under coordinate transformations.

\begin{figure}[h] 
\includegraphics[]{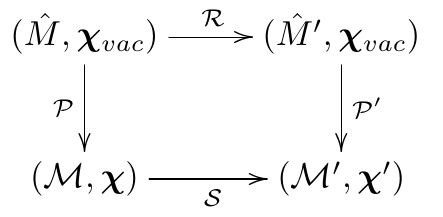}
\caption{Commutative diagram of analog space-times in different coordinates, showing that the analogs $\mc$ and $\mc'$ may be related by $\mathcal{S}$, but $\mathcal{S}$ is not associated to the coordinate transformation $\mathcal{R}$.  The analog model is not covariant under coordinate transformations in $\hat{M}$.}
\label{Fig:CommDiagram}
\end{figure}

On the other hand, since the projection map is freely specifiable, we may find related projections $\mathcal{P}$ and $\mathcal{P}'$ such that the same physical analog is produced, but this is not the same thing as a coordinate transformation in the analog.
Let $\mathcal{R}:\hat{M}\to \hat{M}'$ be a coordinate transformation in $\hat{M}$ with Jacobian matrix $\bm{J}$, and let $\mathcal{S}:\mathcal{M}\to\mathcal{M}'$ be an independent coordinate transformation in Minkowski space-time $\mathcal{M}$ with Jacobian matrix $\bm{L}$, as depicted in Fig.~\ref{Fig:CommDiagram}.
Given the analog projection $\mathcal{P}$, $\mc$ is given by Eq.~(\ref{Eq:chi-general}), which may be related to the primed quantities by
\begin{multline}
\mc\indices{_{\alpha\beta}^{\mu\nu}} = -\left(L\indices{^{\alpha'}_{\alpha}}  L\indices{^{\beta'}_{\beta}} \star\indices{_{\alpha'\beta'}^{\sigma'\rho'}} (L^{-1})\indices{^{\sigma}_{\sigma'}} (L^{-1})\indices{^{\rho}_{\rho'}}\right)  
(\Lambda^{-1})\indices{^A_{\sigma}}(\Lambda^{-1})\indices{^B_{\rho}} \\
\left(J\indices{^{A'}_{A}} J\indices{^{B'}_{B}}\hat{\star}\indices{_{A'B'}^{C'D'}} (J^{-1})\indices{^C_{C'}} (J^{-1})\indices{^D_{D'}} \right) \Lambda\indices{^{\mu}_{C}}\Lambda\indices{^{\nu}_{D}}.
\end{multline}
Inserting two factors of $\bm{L}\bm{L^{-1}}$, this can be written as
\begin{multline}
 \mc\indices{_{\alpha\beta}^{\mu\nu}} = -L\indices{^{\alpha'}_{\alpha}}  L\indices{^{\beta'}_{\beta}} \left( \star\indices{_{\alpha'\beta'}^{\sigma'\rho'}} \Lambda\indices{^{A'}_{\sigma'}}\Lambda\indices{^{B'}_{\rho'}} \hat{\star}\indices{_{A'B'}^{C'D'}} (\Lambda^{-1})\indices{^{\mu'}_{C'}} (\Lambda^{-1})\indices{^{\nu'}_{D'}} \right)  (L^{-1})\indices{^{\mu}_{\mu'}} (L^{-1})\indices{^{\nu}_{\nu'}} \\
  = L\indices{^{\alpha'}_{\alpha}}  L\indices{^{\beta'}_{\beta}} \chi\indices{_{\alpha'\beta'}^{\mu'\nu'}}  (L^{-1})\indices{^{\mu}_{\mu'}} (L^{-1})\indices{^{\nu}_{\nu'}}
\end{multline}
with 
\begin{equation}
 \Lambda\indices{^{\mu'}_{C'}} = L\indices{^{\mu'}_{\mu}}  \Lambda\indices{^{\mu}_{C}} (J^{-1})\indices{^{C}_{C'}}
\end{equation}
or in terms of index-free matrix multiplication
\begin{equation}
\bm{\Lambda'} =  \bm{L}\bm{\Lambda}\bm{J}^{-1}.
\end{equation}
Since $\bm{\Lambda}' = \ed \mathcal{P}'$, this says that by choosing $\mathcal{P}'$ to be the related projection
\begin{equation} \label{Eq:CompositionOfMaps}
 \mathcal{P}' = \mathcal{S} \circ \mathcal{P} \circ \mathcal{R}^{-1}
\end{equation}
that makes Fig.~\ref{Fig:CommDiagram} commutative, $\mc$ and $\mc'$ will represent the same physical analog medium.
However, the coordinates in the analog are still completely independent of the coordinates in $\hat{M}$.
One could choose $S= id$ and we see that the action of $\mathcal{P}'$ is simply to undo the coordinate transformation $\mathcal{R}$, which returns us to the original configuration in the original analog coordinates.

Since the projection is freely specifiable, $\hat{M}$ may be represented by an infinite number of physically inequivalent analog media in Minkowski space-time.
The projection into the analog is not unique and there does not exist a canonical choice of $\mathcal{P}$.
Indeed, as we have seen, a single choice of $\mathcal{P}$ is not even possible if one wants different coordinate descriptions of a vacuum manifold to have the same physical analog representation.
However, this wide freedom in the choice of $\mathcal{P}$ also implies that the analog model only has meaning when interpreted with respect to the chosen map.  
Furthermore, since a curved vacuum space-time cannot be isometrically mapped into a Minkowskian analog, not all features of light propagation in the curved manifold can be simultaneously emulated by the analog.  In subsequent sections it is shown that the choice of projection map should be informed by which feature of light propagation one wishes to emulate.

\subsection{Coordinate identity analogs of FLRW space-time}\label{subsec:PlebanskiAnalog-FLRW}
To illustrate the physical inequivalence of analogs obtained using $\mathcal{P}_{id}$ with different coordinate descriptions of the same space-time, we consider FLRW space-time in conformal, comoving, and physical coordinates, and find that the corresponding analogs are not related by a coordinate transformation.

Friedmann-Lama\^{i}tre-Robertson-Walker space-time is often described by comoving coordinates $(t,r_c,\theta,\phi)$ with the line element 
\begin{equation}\label{Eq:comovFLRW}
ds^2=-dt^2+a(t)^2\left(d{r_c}^2+{r_c}^2d\theta^2+{r_c}^2\sin^2\theta~d\phi^2\right).
\end{equation}
In this form it may be seen that a spatial hypersurface of the FLRW space-time expands or contracts with time through the scale factor $a(t)$ and may be used to model an expanding or contracting universe.
From Eq.~(\ref{Eq:TrivialAnalog}), and Eqs.~(\ref{Eq:G=StarFCompon})-(\ref{Eq:Material}), the corresponding trivial dielectric analog of FLRW space-time in comoving coordinates is given by
\begin{subequations}\label{Eq:comovMat} 
\begin{align}
\bar{\bar\mu}&=\bar{\bar\varepsilon}=a(t)~\mathbb{I}_3,
\label{Eq:comov-mu-epsilon}
\\
^h\bar{\bar{\gamma}}&=^e\bar{\bar{\gamma}}=\mathbf{0}.\label{Eq:comov-gamma}
\end{align}
\end{subequations}
Thus for this coordinate representation, the analog model is an isotropic but time-dependent medium.

Alternatively, FLRW is actually a conformally rescaled form of Minkowski space-time. Defining the conformal time
\begin{equation}\label{Eq:confTime}
\eta=\int_0^t\frac{dt}{a(t)},
\end{equation}
the FLRW space-time now has the line element
\begin{equation}\label{Eq:confFLRW}
ds^2=a(\eta)^2\left(-d\eta^2+d{r_c}^2+{r_c}^2d\theta^2+{r_c}^2\sin^2\theta~d\phi^2\right)
\end{equation}
and we find that the resulting analog is just empty Minkowski space-time
\begin{subequations}\label{Eq:confMat} 
\begin{align}
\bar{\bar\mu}&=\bar{\bar\varepsilon}=\mathbb{I}_3,
\label{Eq:conf-mu-epsilon}
\\
^h\bar{\bar{\gamma}}&=^e\bar{\bar{\gamma}}=\mathbf{0}.\label{Eq:conf-gamma}
\end{align}
\end{subequations}
This result highlights an important feature of dielectric analog space-times in that, as far as the light cone is concerned, a dielectric analog actually represents an equivalence class of conformally related space-times.
The reason is that the $\star$ operators in Eq.~(\ref{Eq:chi-general}) are conformally invariant and so $\mc$ is also conformally invariant.
Thus given a coordinate description that is conformally related to Minkowski space-time we should not be surprised that the analog is just Minkowski space-time. 

In the comoving coordinates introduced above, a spatial hypersurface is seen to expand even though the spatial coordinate points themselves remain fixed.
This implies that the physical distance between two points of fixed comoving coordinates is actually increasing [or decreasing depending on $a(t)$] with time.
The physical distance to an object at fixed comoving coordinate is
\begin{equation}
 r_p =  a(t) r_c.
\end{equation}
An observer would naturally interpret the behavior of objects at fixed comoving coordinates as moving away from each other with recessional velocities that are proportional to their distance - a rule known as Hubble's law.
This Hubble flow may be conveniently described by the physical coordinates $(t,r_p,\theta,\phi)$ with line element
\begin{equation}\label{Eq:physFLRW}
 ds^2=-\left(1-\left(\frac{a'(t)}{a(t)}\right)^2{r_p}^2\right)dt^2-2 \frac{a'(t)}{a(t)}r_p\,dr_p \,dt+d{r_p}^2+{r_p}^2d\theta^2+{r_p}^2\sin^2\theta \,d\phi^2.
\end{equation}
The same procedure as used before gives the physical coordinates analog as
\begin{subequations}\label{Eq:physMat} 
\begin{align}
  \bar{\bar{\mu}}&=\bar{\bar{\varepsilon}}=\left(
\begin{array}{ccc}
 1 & 0 & 0 \\
 0 & \frac{a(t)^2}{a(t)^2-{r_p}^2 a'(t)^2} & 0 \\
 0 & 0 & \frac{a(t)^2}{a(t)^2-{r_p}^2 a'(t)^2} \\
\end{array}
\right),
\label{Eq:phys-mu-epsilon}
\\
^h\bar{\bar{\gamma}}&=\left(^e\bar{\bar{\gamma}}\right)^{T}=\left(
\begin{array}{ccc}
 0 & 0 & 0 \\
 0 & 0 & -\frac{{r_p} a(t) a'(t)}{a(t)^2-{r_p}^2 a'(t)^2} \\
 0 & \frac{{r_p} a(t) a'(t)}{a(t)^2-{r_p}^2 a'(t)^2} & 0 \\
\end{array}
\right),\label{Eq:phys-gamma}
\end{align}
\end{subequations}
which is a complicated anisotropic magnetoelectric medium.

In each of these cases, the target manifold has been specified as Minkowski space-time with spherical coordinates.  
It is therefore clear that Plebanski's approach with analog projection map $\mathcal{P}_{id}$ results in different physical analog models for different coordinate descriptions of the same vacuum space-time.

\section{Light Propagation in FLRW Analogs in Comoving Coordinates}\label{sec:LightPropagation}

From the time of Eddington \cite{Eddington1920} and Gordon \cite{Gordon1923}, through Plebanski \cite{Plebanski1960pr} and to the present, the study of dielectric analog space-times has been largely focused on  coordinate descriptions of the geometric optics ray trajectories of light through the analog, and it appears to have been widely assumed that this coordinate description suffices as a faithful representation of light propagation through the curved space-time $\hat{M}$.
However, a coordinate description of ray trajectories in the analog is of limited value because the coordinates in $\hat{M}$ and the analog coordinates refer to different metrics. 
So even if rays in the analog pass through points with the same coordinate labels as rays in $\hat{M}$, they cannot simultaneously replicate all kinematical aspects of light propagation in $\hat{M}$, such as the fractional expansion, shear, and vorticity of a congruence;
quantities that carry characteristic information about the space-time.  For example, the fractional expansion of a null congruence tells us about the presence of singularities, trapped surfaces, and horizons, and so would need to be properly analyzed for any analog that claimed to replicate such features.

In fact, it turns out that the ability of an analog model to faithfully emulate these characteristic features is quite limited.
We focus on the fractional expansion of a congruence of light rays in the analog, and find that while a projection may be found such that the expansion of a congruence in the analog matches the expansion of a congruence in $\hat{M}$, the required map is specific to the chosen congruence; any other congruence in the analog will fail to have the correct expansion.

The discussion of light rays falls within the realm of geometric optics.
To arrive at the geometric optics limit in the analog, one may begin by assuming a solution for the 4-vector potential $A_{\mu} = \tilde{A}_{\mu} e^{\frac{i}{\lambda}S}$ and $\mF = \ed \mA$.
Inserting this into Maxwell's equations (\ref{Eq:Hom}) and (\ref{Eq:InHom}) and taking the stationary phase approximation $\lambda\to 0$ leads to \cite{Thompson2012aiep}
\begin{equation}\label{Eq:eikonal}
X\indices{_\alpha^\rho}\tilde{A}_\rho=0,
\end{equation}
where
\begin{equation}\label{Eq:X}
X\indices{_\alpha^\rho}=\eta^{\mu\nu}~\chi\indices{_{\nu\alpha}^{\sigma\rho}}k_\mu~k_\sigma,
\end{equation}
$\eta^{\mu\nu}$ is the background Minkowski metric, and $k_{\mu} = \partial_{\mu}S$ is the wave vector. 
The existence of nontrivial solutions to Eq.~(\ref{Eq:eikonal}) requires $\det(X\indices{_{\alpha}^{\rho}})=0$.
It turns out that this determinant condition is satisfied identically, so the subsidiary condition
\begin{equation}\label{Eq:H-J-general}
\mathcal{H}\left(x^\alpha,k_\beta\right)=\textrm{det}\left(X\indices{^i_j}\right)=0,
\end{equation}
must be satisfied, where the indices $i,j$ only run over the purely spatial components. 

For impedance-matched media such as that obtained by projection from a vacuum space-time $\hat{M}$, $\mathcal{H}$ will always be of the form
\begin{equation}
\mathcal{H}(x^{\alpha},k_{\beta}) = B(x^{\alpha}){k_0}^2 H^2
\end{equation}
where $H$ can be thought of as a second-order polynomial in $k_0$, and $B(x^{\alpha})$ is some factor that does not depend on $k_{\beta}$.
The condition $\mathcal{H}=0$ will be everywhere satisfied on a solution curve if and only if $H=0$, so the condition $\mathcal{H}=0$ becomes the condition $H=0$.
For impedance-matched media, $H$ can always be written as
\begin{equation}
H = \frac12 \gamma^{\mu\nu} k_{\mu}k_{\nu},
\end{equation}
which is of the same form $H$ assumes in vacuum space-times, but where $\gamma^{\mu\nu}$ would be replaced with the space-time metric $g^{\mu\nu}$. Thus, in media the tensor $\gamma^{\mu\nu}$ is often referred to as the ``optical metric,'' but should be understood as containing information on both the medium parameters and background metric.
Since $H(x^{\alpha},k_{\beta})$ is a function on the cotangent bundle that vanishes everywhere on a solution curve, then the exterior derivative of $H$ also vanishes, $\ed H = 0$, leading to Hamilton's canonical equations
\begin{equation} \label{Eq:Hamilton}
 \begin{aligned} 
 \dot{x}^{\alpha} & = \frac{\partial H}{\partial k_{\alpha}}, \\
 \dot{k}_{\alpha} & = - \frac{\partial H}{\partial x^{\alpha}},
 \end{aligned}
\end{equation}
which may be integrated to find the ray trajectories.

\subsection{Particle horizon and redshift in the trivial FLRW analog}
Before showing the failure of the analog to correctly replicate kinematical aspects of light propagation in $\hat{M}$, we first examine some examples where an analog generated by $\mathcal{P}_{id}$ has potentially interesting properties.
Consider the projection $\mathcal{P}_{id}$ between FLRW space-time in comoving coordinates and the spherical coordinates of Minkowski space-time.
The optical metric is $\gamma^{\mu\nu}=g^{\mu\nu}$, so one might naively expect light propagation in the analog to fully mimic light propagation in the vacuum.

The wave vector for a future-directed radially propagating ray in the analog is
\begin{equation} \label{Eq:momentum}
 k_{\mu} = (-k_0(t,r),k_1(t,r),0,0).
\end{equation}
From
\begin{equation}\label{Eq:Cond1}
H=\frac{1}{2} \left(\frac{{k_1}^2}{a(t)^2}-{k_0}^2\right) = 0
\end{equation}
we get
\begin{equation}\label{Eq:Cond2}
k_1=\pm a(t) k_0,
\end{equation}
in which the positive (negative) sign corresponds to outgoing (ingoing) rays.
The tangent to the curve is found from the first of Hamilton's equations (\ref{Eq:Hamilton}) to be
\begin{equation}\label{Eq:v-analog}
u^\alpha=\left(k_0,\pm\frac{k_0}{a(t)},0,0\right).
\end{equation}
Meanwhile, the second of Hamilton's equations (\ref{Eq:Hamilton}) provides an evolution equation for $k_{\mu}$
\begin{equation}\label{Eq:evolution-kdot}
\dot k_\alpha=\left(\frac{{k_0}^2  a'(t)}{a(t)}, 0, 0, 0\right).
\end{equation}
Thus we have a single equation for $k_0$,
\begin{equation}
-\frac{\partial}{\partial\tau}\ln(k_0)=k_0 \frac{\partial}{\partial t}\ln(a(t)) = k_0 \left(\frac{dt}{d\tau}\right)^{-1}\frac{\partial}{\partial\tau} \ln(a(t(\tau))).
\end{equation}
But from the first of Hamilton's equations we know that $\frac{dt}{d\tau }=u^0=k_0$.
The solution for $k_0$ is straightforwardly
\begin{equation}
k_0=\frac{{a_0} k_{0i}}{a(t)},
\end{equation}
and substitution into Eq.~(\ref{Eq:v-analog}) yields
\begin{equation}\label{Eq:v-analog-new}
u^\alpha=\frac{a_0 k_{0i}}{a(t)}\left(1,\pm\frac{1}{a(t)},0,0\right).
\end{equation}
For an outgoing ray, one deduces
\begin{equation}\label{Eq:dr/dt}
\frac{dr}{dt}=\frac{u^r}{u^t}=\frac{1}{a(t)}.
\end{equation}

If light commences propagating radially inward from all points in the analog at some time $t_0$, then the largest radius in the analog from which an observer at $r=0$ has received a signal at time $t$ is
\begin{equation}\label{Eq:PH}
\int^0_{r_{PH}}dr=\int_{t_0}^{t}-\frac{dt}{a(t)}.
\end{equation}
The above relation replicates the most common definition of the particle horizon - the most distant point in the universe from which we have received light since the big bang. Furthermore, if we consider the frequency of two successive pulses, separated temporally by an interval $\Delta t$, propagating in the analog and reaching an observer, then Eq.~(\ref{Eq:dr/dt}) gives the frequency shift between the two pulses by
\begin{equation}\label{Eq:redshift}
z+1=\frac{\omega_0}{\omega(t)}.
\end{equation}
For a detailed derivation of frequency shift and particle horizon from Eq.~(\ref{Eq:dr/dt}), see for example section 10 of Ref.~\cite{Ryder2009}.  
Note that redshifts in the analog are facilitated by the time dependence of the medium parameters \cite{Cummer2011jo}.
The above results show that at least some of the basic and well-known effects observed in FLRW space-time can be replicated in the analog.

Before moving on, it is worth pointing out some generic features of light propagation in the analog.
First, for both ingoing and outgoing rays $u^\alpha u_\alpha<0$, indicating that light rays in the analog are time-like.
Second, $\nabla_{\textbf{u}}\textbf{u}\neq0$ shows that the light rays are not geodesic with respect to the background Minkowski space-time, nor is it possible to find a reparametrization of the curve so as to make it geodesic.
Third, $u^{\alpha}k_{\alpha}=0$ is valid in the analog, but it is not true that $\eta_{\alpha\beta}u^{\beta} = k_{\alpha}$.

\subsection{Congruence expansion}\label{subsec:congruenceExpansion}
Despite the fact that the projection $\mathcal{P}_{id}$ results in an optical metric $\gamma^{\mu\nu} = g^{\mu\nu}$, light propagation in the analog does not faithfully replicate all aspects of light propagation in the vacuum.
The reason is that although the optical metric is identical to the vacuum metric, most physical measures of light propagation in the analog are actually made with respect to the background Minkowski metric, such as the kinematical decomposition of a congruence.
This decomposition provides physically meaningful measures such as the fractional rate of expansion of the cross section of a laser beam, and in curved space-times is instrumental in determining the existence of trapped surfaces and horizons.
As such, we focus on the fractional rate of expansion of a radial congruence of light propagating through the analog, and find that projecting with $\mathcal{P}_{id}$ does not reproduce the vacuum rate of expansion.
Furthermore, we find that although it is possible to construct a projection tailored such that the expansion is preserved, it can only be preserved for a single choice of congruence at a time.
Preserving the expansion for a different choice of congruence requires a different projection.

\subsubsection{Expansion of a radial congruence in FLRW}
Before analyzing the expansion of a congruence in the analog, we first find the expansion in the vacuum space-time that we are attempting to emulate.
We again consider future-directed radially propagating rays.  
In the vacuum, these rays are null geodesics for which $u^{A} k_{A}=0$ and $k_{A}=g_{AB}u^{B}$.
It may be shown that the evolution of a congruence centered on the curve $u^{A}$ is governed by $u_{A;B}$ \cite{Poisson}.
The expansion is the fractional rate of change of the transverse cross section of the congruence, which is equal to the trace of the projection of $u_{A;B}$ into the transverse subspace.
The light cone at every point is spanned by both outgoing and ingoing rays, meaning that the transverse subspace is only two dimensional.
One may find a projection operator from the inherited metric on the transverse subspace \cite{Poisson,Faroni2015}
\begin{equation}\label{Eq:2-metric-null}
h^{AB}=g^{AB}+ u^A v^B + u^B v^A.
\end{equation}
In the literature, $v^{A}$ is typically introduced as an ``auxiliary'' null vector such that $u^A v_A=-1$.
The expansion of the vacuum congruence about $u^{A}$ is
\begin{equation}\label{Eq:expansion-null-general}
\Theta_u=h^{AB}u_{A;B}
\end{equation}
from which we find the expansion of radially outgoing and ingoing rays to be, respectively,
\begin{subequations}\label{Eq:expansion-null-general-FRW}
\begin{align}
\Theta_u=\frac{2 {a_0} {k_{0i}} \left(r a'(t)+1\right)}{r a(t)^2},\label{Eq:expansion-null-general-FRW-a}\\
\Theta_v=\frac{2 {a_0} {k_{0i}} \left(r a'(t)-1\right)}{r a(t)^2}.\label{Eq:expansion-null-general-FRW-b}
\end{align}
\end{subequations}

\subsubsection{Expansion of a radial congruence in the analog}

By contrast, we have said that rays in the analog follow nongeodesic time-like curves with respect to the background metric, so 
it is tempting to examine the kinematics of light in media with the formalism for time-like congruences \cite{Poisson}.
However, with the advent of metamaterials that offer a great degree of control over light propagation, we find the need for an analysis that can track a congruence that smoothly varies from null to time-like, and even to space-like.
The light cone at every point is still defined by a set of two linearly independent rays - left/right or ingoing/outgoing rays.
Thus the transverse subspace is still two dimensional, as would be expected for say, the cross section of a laser beam.
 
The projection operator defined by Eq.~(\ref{Eq:2-metric-null}) made use of the fact that $k_{\mu}=g_{\mu\nu}u^{\nu}$ in the vacuum, a relation that does not hold in the analog.
It is therefore necessary to be cognizant of index placement in the analog, and to construct a more general projection operator.  
It was previously mentioned that $H$ could be considered as a second-order polynomial in $k_{\mu}$.
Thus at every point we actually have two solutions, corresponding to ingoing and outgoing rays.  
Let the pairs $(u^\alpha, k_\alpha)$ and $(v^\alpha, \ell_\alpha)$ denote the tangent vector and wave covector for outgoing and ingoing rays, respectively, which satisfy
\begin{subequations}\label{Eq:conditions-on-the-pairs}
\begin{align}
u^\alpha k_\alpha=0,\quad\quad\quad u^\alpha\ell_\alpha\neq0,\label{Eq:conditions-on-the-pairs-u}\\
v^\alpha \ell_\alpha=0,\quad\quad\quad v^\alpha k_\alpha\neq0.\label{Eq:conditions-on-the-pairs-v}
\end{align}
\end{subequations}

With these fields, the projection operator generalizes to \cite{Thompson2016}
\begin{equation}\label{Eq:projection-operator}
h\indices{_\beta^\alpha}=\delta_\beta^\alpha-\frac{\ell_\beta u^\alpha}{\ell_\sigma u^\sigma}-\frac{k_\beta v^\alpha}{k_\sigma v^\sigma}.
\end{equation} 
One may readily verify that $u^\beta h\indices{_\beta^\alpha}=v^\beta h\indices{_\beta^\alpha}=h\indices{_\beta^\alpha}k_\alpha
=h\indices{_\beta^\alpha}\ell_\alpha=0$, and also that $h\indices{_\mu^\beta}h\indices{_\beta^\alpha}=h\indices{_\mu^\alpha}$.
Since $h\indices{_\beta^\alpha}$ is orthogonal to both $u^{\beta}$ and $v^{\beta}$ the transverse subspace is still only two dimensional, which is confirmed by the trace $h\indices{_\alpha^\alpha}=2$.
This construction of $h\indices{_\beta^\alpha}$ follows directly from Maxwell's equations in media and is equally applicable to congruences that are null, time-like, or space-like with respect to the background metric.

Using this projection operator to calculate the transverse fractional expansion of a congruence in the analog with $u^{\alpha}$ and $v^{\alpha}$ as in Eq.~(\ref{Eq:v-analog-new}), one finds 
\begin{equation}\label{Eq:expansion-trivial-analog}
\Theta_u=-\Theta_v = h\indices{_\beta^\alpha} u\indices{^{\beta}_{;\alpha}} = \frac{2 {a_0} {k_{0i}}}{r a(t)^2},
\end{equation}
which disagrees with the vacuum expansion found in Eqs.~(\ref{Eq:expansion-null-general-FRW-a}) and (\ref{Eq:expansion-null-general-FRW-b}).

\subsection{Expansion-replicating analog model}\label{subsec:Constraints}
We have found that an analog model based on $\mathcal{P}_{id}$ fails to correctly mimic the fractional rate of expansion of the congruence despite the fact that the optical metric and the coordinate description of light curves are the same as in the original space-time.
Since the projection is freely specifiable, it is natural to ask whether a different choice of projection could reproduce the correct expansion rate at the expense of the optical metric and coordinate description of curves.

The quantity $B\indices{^\alpha_\beta}=u\indices{^\alpha_{;\beta}}$ governs the evolution of the separation between the curve with tangent $u^{\alpha}$ and a nearby curve in the congruence \cite{Poisson}.
Let 
\begin{equation}\label{Eq:B-transverse}
\bar{B}\indices{^\alpha_\beta}=h\indices{_\mu^\alpha} h\indices{_\beta^\nu}u\indices{^\mu_{;\nu}}
\end{equation} 
be the projection of $B\indices{^\alpha_\beta}$ into the transverse subspace, thereby describing the purely transverse evolution of the congruence.
The fractional expansion is the trace of $\bar{B}\indices{^\alpha_\beta}$, so by calculating $\bar{B}\indices{^\alpha_\beta}$ in both the vacuum and the analog we may find a condition that determines an expansion-preserving projection.

Expanding the covariant derivative, we have 
\begin{equation}\label{Eq:B-transverse-vacuum}
\bar{B}\indices{^A_B}=h\indices{_M^A}h\indices{_B^N}\left(u\indices{^M_{,N}}+\Gamma\indices{^M_{NC}}u^C\right),
\end{equation}
in vacuum, and
\begin{equation}\label{Eq:B-transverse-analog}
\bar{B}\indices{^\alpha_\beta}=h\indices{_\mu^\alpha}h\indices{_\beta^\nu}\left(u\indices{^\mu_{,\nu}}+\Gamma\indices{^\mu_{\nu\gamma}}u^\gamma\right),
\end{equation}
in the analog, where capital latin indices refer to the vacuum space-time $\hat{M}$, lowercase greek indices refer to the analog in Minkowski space-time $\mathcal{M}$, while $\Gamma\indices{^M_{NC}}$ and $\Gamma\indices{^\mu_{\nu\gamma}}$ are the Christoffel symbols for $\hat{M}$ and $\mathcal{M}$, respectively. 
Let $\mathcal{P}: \hat{M}\to\tilde{M}\subseteq\mathcal{M}$ be the sought-after analog projection map, with associated Jacobian $\mlambda=\ed \mathcal{P}$ .
The ingoing and outgoing rays in the analog are related to those in $\hat{M}$ by
\begin{subequations}\label{Eq:transform-vacumm-to-analog-u,v}
\begin{align}
u^\alpha=\Lambda\indices{^\alpha_A}u^A ,\quad\quad\quad \ell_\alpha=\left(\Lambda^{-1}\right)\indices{^A_\alpha}\ell_A,\label{Eq:transform-vacumm-to-analog-u}\\
v^\alpha=\Lambda\indices{^\alpha_A}v^A ,\quad\quad\quad k_\alpha=\left(\Lambda^{-1}\right)\indices{^A_\alpha}k_A.\label{Eq:transform-vacumm-to-analog-v}
\end{align}
\end{subequations}
from which it follows that the projection operator transforms as
\begin{equation}\label{Eq:transform-vacumm-to-analog-h}
h\indices{_\mu^\alpha}=\Lambda\indices{^\alpha_A}\left(\Lambda^{-1}\right)\indices{^M_\mu}
h\indices{_M^A}
\end{equation}
and  
\begin{equation}\label{Eq:transform-vacumm-to-analog-differentiation-of-u}
u\indices{^\mu_{,\nu}}=\Lambda\indices{^\mu_M}\left(\Lambda^{-1}\right)\indices{^N_\nu}u\indices{^M_{,N}}
+\left(\left(\Lambda^{-1}\right)\indices{^N_\nu}\partial_N\Lambda\indices{^\mu_M}\right)u^M.
\end{equation}
Finally, one can recast the analog $\bar B\indices{^\alpha_\beta}$ in Eq.~(\ref{Eq:B-transverse-analog}) in terms of quantities in $\hat{M}$ as \cite{Thompson2016}
\begin{equation}\label{Eq:transform-vacumm-to-analog-B-analog}
\bar B\indices{^\alpha_\beta}=\Lambda\indices{^\alpha_A}\left(\Lambda^{-1}\right)
\indices{^B_\beta}h\indices{_M^A}h\indices{_B^N}
\left[u\indices{^M_{,N}}+\tensor[^*]{\Gamma}{^M_{NS}}u^S\right],
\end{equation}
with
\begin{equation}\label{Eq:GammaStar}
{^*}\Gamma\indices{^M_{NS}}=\left(\Lambda^{-1}\right)\indices{^M_\mu}\Lambda\indices{^\nu_N}
\Lambda\indices{^\gamma_S}{^a}\Gamma\indices{^\mu_{\nu\gamma}}
+\left(\Lambda^{-1}\right)\indices{^M_\mu}\partial_N\Lambda\indices{^\mu_S}.
\end{equation}

The above relations show that $\bar B\indices{^\alpha_\beta}$ is not quite the same as a simple mapping of $\bar B\indices{^A_B}$ from the vacuum to the analog since ${^*}\Gamma\indices{^M_{NS}}$ is not necessarily the same as $\Gamma\indices{^M_{NS}}$. 
Taking the trace of Eqs.~(\ref{Eq:B-transverse-vacuum}) and (\ref{Eq:transform-vacumm-to-analog-B-analog}) we are able to compare ${^v}\Theta=\bar B\indices{^A_A}$ and ${^a}\Theta=\bar B\indices{^\alpha_\alpha}$ as quantities in $\hat{M}$. 
One finds
\begin{subequations}\label{Eq:traces-of-B-vacuum,analog}
\begin{align}
{^v}\Theta=\bar B\indices{^A_A}=h\indices{_M^N}\left[u\indices{^M_{,N}}+\Gamma\indices{^M_{NS}}u^S\right],
\label{Eq:traces-of-B-vacuum}\\
{^a}\Theta=\bar B\indices{^\alpha_\alpha}=h\indices{_M^N}\left[u\indices{^M_{,N}}+{^*}\Gamma\indices{^M_{NS}}u^S\right].
\label{Eq:traces-of-B-analog}
\end{align}
\end{subequations}
Hence, in order to have an analog that correctly replicates the expansion of a congruence ${^a}\Theta={^v}\Theta$, we must satisfy the condition
 \begin{equation}\label{Eq:constraint-on-the-analog}
h\indices{_M^N}\left[{^*}\Gamma\indices{^M_{NS}}-\Gamma\indices{^M_{NS}}\right]u^S=0.
\end{equation}
Now, if one wished to have an analog that correctly replicates the expansion for all congruences then one would have to satisfy the stronger condition
\begin{equation}
{^*}\Gamma\indices{^M_{NS}} = \Gamma\indices{^M_{NS}},
\end{equation}
which says that $\mathcal{P}$ pulls $\Gamma\indices{^{\mu}_{\nu\gamma}}$ from the Minkowski background of the analog into $\Gamma\indices{^M_{NS}}$.
However, by Gauss' Theorema Egregium such a map cannot exist because the curved space-time $\hat{M}$ and the flat Minkowski background of the analog are not isometric.
Therefore, we have found that a single map cannot satisfy Eq.~(\ref{Eq:constraint-on-the-analog}) for all congruences.  The best one can hope for is to satisfy Eq.~(\ref{Eq:constraint-on-the-analog}) for a given choice of congruence.

As an example, we seek the map that will preserve the expansion of the radially outgoing congruence with tangent vector given by Eq.~(\ref{Eq:v-analog-new}), e.g.\
\begin{subequations}\label{Eq:arbitrary-congruence}
\begin{align}
u^A=\left(\frac{{a_0} {k_{0i}}}{a(t)},\frac{{a_0} {k_{0i}}}{a(t)^2},0,0\right),\quad\quad\quad k_A=g_{AB}u^B,
\label{Eq:arbitrary-congruence-u}\\
v^A=\left(\frac{{a_0} {k_{0i}}}{a(t)},-\frac{{a_0} {k_{0i}}}{a(t)^2},0,0\right),\quad\quad\quad \ell_A=g_{AB}v^B.
\label{Eq:arbitrary-congruence-v}
\end{align}
\end{subequations}
Given the symmetries of the space-time and the congruence, we suspect that the desired projection from FLRW to the analog has the form
\begin{equation}\label{Eq:analog-map-choice}
\left(t_{\mathcal{M}},r_{\mathcal{M}},\theta_{\mathcal{M}},\phi_{\mathcal{M}}\right) = \mathcal{P}\left(t_{\hat{M}}, r_{\hat{M}}, \theta_{\hat{M}}, \phi_{\hat{M}}\right)=\left(t_{\hat{M}}, f(t_{\hat{M}})r_{\hat{M}}, \theta_{\hat{M}}, \phi_{\hat{M}}\right),
\end{equation}
with the associated Jacobian
\begin{equation}\label{Eq:analog-map-choice-Jacobian}
\Lambda\indices{^\alpha_A}=\left(%
\begin{array}{cccc}
  1 & 0 & 0 & 0 \\
  f'(t) r_{\hat{M}} & f(t) & 0 & 0 \\
  0 & 0 & 1 & 0\\
 0 & 0 & 0 & 1 \\
\end{array}%
\right).
\end{equation}
Since $t_{\mathcal{M}} = t_{\hat{M}}$ we denote both by $t$. 
Equation~(\ref{Eq:constraint-on-the-analog}) provides the differential equation
\begin{equation}\label{Eq:equation-for-f(t)}
\frac{2 {a_0} {k_{0i}} \left(a(t) f'(t)-f(t) a'(t)\right)}{a(t)^2 f(t)}=0,
\end{equation}
which has the solution
\begin{equation}\label{Eq:solution-for-f(t)}
f(t)=c_1 a(t),
\end{equation}
where $c_1$ is an integration constant. 
Thus, for the particular choice of congruence made above we have found an expansion-preserving map.  For any other choice of congruence we would have to go through the same procedure, for which the outcome would be the specification of a totally different analog projection map.

\section{Dielectric analog space-times and transformation optics}\label{sec:Misconceptions}
One of the pioneering formulations of transformation optics was based on Eqs.~(\ref{Eq:Plebanskiepsilon}) and (\ref{Eq:magnetoelectric}); in other words, on Plebanski's analog space-times model using $\mathcal{P}_{id}$ as the projection \cite{Leonhardt2006njp1}.
However, we argue that, while closely related, analog space-times are not the same thing as TO.
The typical explanation of TO is that one starts in Minkowski space-time and performs a coordinate transformation which results in a new metric describing a curved space-time; this curved space-time metric  is then matched to a dielectric medium via Eqs.~(\ref{Eq:Plebanskiepsilon}) and (\ref{Eq:magnetoelectric}).
This explanation is problematic because the space-time curvature is a diffeomorphism invariant, so it is not possible to use a coordinate transformation to generate a curved space-time and associated metric.
One could instead argue that the mechanism being employed is a map into a curved space-time that has the desired properties, and a subsequent projection into an analog medium.
Such an argument actually bypasses the initial coordinate transformation because it implies that there exist curved vacuum manifolds that possess all possible variations of light propagation, and it is merely a matter of identifying the desired vacuum manifold and projecting to an analog.
But we have just shown that analog space-times suffer from three serious issues that make them a poor choice as the basis for TO.
\begin{enumerate}
 \item There is no canonical choice of projection map, so any formulation of TO based on this approach is not unique.
 \item Analog space-times are not covariant under coordinate transformations, while any useful formulation of TO should be covariant.
 \item Most concerningly, analog space-times do not fully replicate the behavior of light propagation in the vacuum.  Thus any \textit{transformation medium} obtained as the analog of a space-time with the desired light propagation would simply fail to work as expected.
\end{enumerate}

\begin{figure}[ht]
 \includegraphics[]{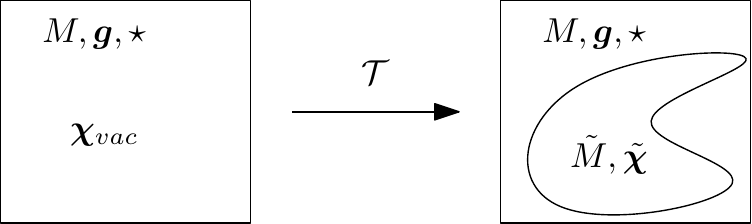}
 \caption{In transformation optics, the space-time manifold, metric, and coordinates all remain unchanged under the corporeal transformation $\mathcal{T}$.  An initial distribution of electromagnetic fields on $M$ are transformed by the pullback of $\mathcal{T}^{-1}$, and only have support in $\tilde{M}=\mathcal{T}(M)\subseteq M$.}
 \label{Fig:TO}
\end{figure}

Instead, it has been shown that a more consistent and fully covariant formulation of TO can be obtained through what has been termed a \textit{corporeal transformation} \cite{Thompson2011jo1,Thompson2011jo2,Thompson2015pra}
\begin{equation}
 \mathcal{T}: M \to \tilde{M}\subseteq M
\end{equation}
as depicted in Fig.~\ref{Fig:TO}.
The key difference is that we begin and end in the same space-time manifold rather than projecting from one manifold to a different manifold.
The dielectric medium $\tilde{\mc}$ can be solved for in exactly the same manner as finding an analog space-time, but Eq.~(\ref{Eq:chi-general}) becomes instead
\begin{equation}\label{Eq:chi-TO}
  {\tilde{\chi}}\indices{_{\eta\tau}^{\pi\theta}}(x)
       =-\star\indices{_{\eta\tau}^{\lambda\kappa}}|_x~\left(\Lambda^{-1}\right)\indices{^{\alpha}_{\lambda}}|_x~\left(\Lambda^{-1}\right)\indices{^{\beta}_{\kappa}}|_x~\star\indices{_{\alpha\beta}^{\mu\nu}}|_{\mathcal{T}^{-1}(x)}~\Lambda\indices{^{\pi}_{\mu}}|_x~\Lambda\indices{^{\theta}_{\nu}}|_x
\end{equation}
where the only change is that everything refers to the same coordinate system on the same manifold $M$.

Physically this makes more sense. 
For all intents and purposes we cannot engineer the space-time; instead, the space-time remains fixed and we simply introduce the presence of dielectric media to manipulate the fields, an operation described by the corporeal transformation.
The corporeal transformation is also not a coordinate transformation since the introduction of media should not have anything to do with coordinates; thus this basis for TO is fully covariant under coordinate transformations \cite{Thompson2015pra}.

\begin{figure}[ht]
\includegraphics[]{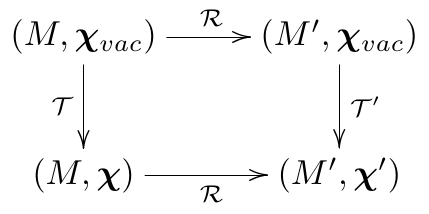} 
\caption{In transformation optics, a coordinate transformation $\mathcal{R}$ in the vacuum configuration corresponds to the same coordinate transformation in the transformation media $\mc$.  Thus TO is fully covariant under coordinate transformations.}
\label{Fig:TOCommDiagram}
\end{figure}

This formulation resolves the three major issues enumerated above.
First, since there is no additional projection map involved, the transformation medium is unique for a given corporeal transformation.
Second, this formulation of TO is fully covariant under coordinate transformations, as depicted in Fig.~\ref{Fig:TOCommDiagram}.
In particular, given a coordinate transformation $\mathcal{R}$ with Jacobian matrix $\bm{J}$ in the vacuum, the corporeal transformation $\mathcal{T}$, metric, and Hodge dual can all be reexpressed in the new coordinates as $\mathcal{T}'$, $\mg'$, and $\star'$, and it follows that
\begin{multline}
 \begin{aligned}
   {\tilde{\chi}}\indices{_{\eta\tau}^{\pi\theta}}(x)
        &=-\left(J\indices{^{\eta'}_{\eta}}J\indices{^{\tau'}_{\tau}}\star\indices{_{\eta'\tau'}^{\lambda'\kappa'}} \left(J^{-1}\right)\indices{^{\lambda}_{\lambda'}}\left(J^{-1}\right)\indices{^{\kappa}_{\kappa'}}\right)
        \left(J\indices{^{\gamma'}_{\lambda}}\left(\Lambda^{-1}\right)\indices{^{\alpha'}_{\gamma'}}\left(J^{-1}\right)\indices{^{\alpha}_{\alpha'}}\right) 
        \left(J\indices{^{\delta'}_{\kappa}}\left(\Lambda^{-1}\right)\indices{^{\beta'}_{\delta'}}\left(J^{-1}\right)\indices{^{\beta}_{\beta'}}\right) \\
        & \qquad   \left(J\indices{^{\sigma'}_{\alpha}}J\indices{^{\rho'}_{\beta}}\star\indices{_{\sigma'\rho'}^{\mu'\nu'}} \left(J^{-1}\right)\indices{^{\mu}_{\mu'}}\left(J^{-1}\right)\indices{^{\nu}_{\nu'}}\right)
        \left(J\indices{^{\psi'}_{\mu}} \Lambda\indices{^{\pi'}_{\psi'}} (J^{-1})\indices{^{\pi}_{\pi'}} \right) 
        \left(J\indices{^{\omega'}_{\nu}} \Lambda\indices{^{\theta'}_{\omega'}} (J^{-1})\indices{^{\theta}_{\theta'}} \right) \\
        & = - J\indices{^{\eta'}_{\eta}}J\indices{^{\tau'}_{\tau}} \star\indices{_{\eta'\tau'}^{\lambda'\kappa'}} (\Lambda^{-1})\indices{^{\alpha'}_{\lambda'}}(\Lambda^{-1})\indices{^{\beta'}_{\kappa'}}\star\indices{_{\alpha'\beta'}^{\mu'\nu'}}\Lambda\indices{^{\pi'}_{\mu'}}\Lambda\indices{^{\theta'}_{\nu'}} (J^{-1})\indices{^{\pi}_{\pi'}}(J^{-1})\indices{^{\theta}_{\theta'}}\\
        & = J\indices{^{\eta'}_{\eta}}J\indices{^{\tau'}_{\tau}} \tilde{\chi}\indices{_{\eta'\tau'}^{\pi'\theta'}}(J^{-1})\indices{^{\pi}_{\pi'}}(J^{-1})\indices{^{\theta}_{\theta'}}.
 \end{aligned}
\end{multline}
So unlike the case for analog space-times, a coordinate change in the initial TO vacuum corresponds to the same coordinate change in the transformation medium.
Third, since TO only refers to a single space-time there is no question about which metric kinematical aspects of light propagation are referred to.
The kinematics of a congruence in the transformation medium follows exactly from the corporeal transformation and is not compromised by an intermediary step in a nonphysical curved space-time.

Two final remarks about the formulation of TO discussed here are that the initial configuration need not be vacuum \cite{Thompson2010pra}, and that it allows one to do TO in curved space-times, not just in Minkowski space-time \cite{Thompson2012jo1,Thompson2015pra}.

\section{Conclusion}\label{sec:conclusion}

By examining the kinematical properties of dielectric analog space-times, we have shown that the projection of a curved space-time $\hat{M}$ into a dielectric analog residing in a space-time of different curvature, such as Minkowski space-time, is limited in its ability to faithfully represent all aspects of light propagation in $\hat{M}$.
This is reminiscent of the challenge faced by cartographers when representing a curved surface on a flat sheet, and follows directly from Gauss' Theorema Egregium which forbids distortion-free mappings between nonisometric spaces.

Although such mappings are well defined, with tensorial field quantities mapped with well-defined differential geometry operations, analog space-times are found to be noncovariant under coordinate transformation in the sense that the analog model is coordinate dependent.  
Two analogs representing two different coordinate descriptions of the same space-time are not necessarily the same physical device described in concomitantly related coordinates.
Relying solely on the Plebanski model of coordinate identification projection $\mathcal{P}_{id}$ would result in two completely different physical devices representing the same space-time.

This helps to highlight the fact that there is no canonical choice of projection map $\mathcal{P}$; an infinite number of projections are possible, and which projection one chooses to employ must be guided by which features of vacuum light propagation one wishes to exhibit in the analog.
This is nicely illustrated by looking at an analog of FLRW in different guises.
It was found that Plebanski's model of FLRW could reproduce some notion of the particle horizon and redshift, but completely failed to adequately represent the kinematical properties of a congruence of light rays, in particular the expansion of a transverse cross section of the congruence.
It is possible to construct a projection map $\mathcal{P}$ such that the expansion is preserved for a given choice of congruence at the expense of distortions in the optical metric and coordinate description of light rays, but not for all congruences simultaneously.

Last, the lack of canonical choice of projection, issues with covariance under coordinate transformations, and the simple fact that the analog does not fully represent light propagation in the vacuum, all cast doubt on the viability of the dielectric analog space-times approach as a basis for transformation optics.
Instead, these issues can be simultaneously resolved by simply assuming that TO is formulated entirely as corporeal transformations on a single manifold.  Such transformations do not touch the metric or the coordinates and only act to actively modify fields; these have a straightforward interpretation as the insertion of a field-modifying medium.

\subsection*{Acknowledgments}
The authors are grateful to Prof.\ J\"{o}rg Frauendiener (U.\ Otago) for useful discussions. R.T.T. was supported by the Royal Society of New Zealand through Marsden Fund Fast Start Grant No. UOO1219.


\begin{thebibliography}{45}%
\makeatletter
\providecommand \@ifxundefined [1]{%
 \@ifx{#1\undefined}
}%
\providecommand \@ifnum [1]{%
 \ifnum #1\expandafter \@firstoftwo
 \else \expandafter \@secondoftwo
 \fi
}%
\providecommand \@ifx [1]{%
 \ifx #1\expandafter \@firstoftwo
 \else \expandafter \@secondoftwo
 \fi
}%
\providecommand \natexlab [1]{#1}%
\providecommand \enquote  [1]{``#1''}%
\providecommand \bibnamefont  [1]{#1}%
\providecommand \bibfnamefont [1]{#1}%
\providecommand \citenamefont [1]{#1}%
\providecommand \href@noop [0]{\@secondoftwo}%
\providecommand \href [0]{\begingroup \@sanitize@url \@href}%
\providecommand \@href[1]{\@@startlink{#1}\@@href}%
\providecommand \@@href[1]{\endgroup#1\@@endlink}%
\providecommand \@sanitize@url [0]{\catcode `\\12\catcode `\$12\catcode
  `\&12\catcode `\#12\catcode `\^12\catcode `\_12\catcode `\%12\relax}%
\providecommand \@@startlink[1]{}%
\providecommand \@@endlink[0]{}%
\providecommand \url  [0]{\begingroup\@sanitize@url \@url }%
\providecommand \@url [1]{\endgroup\@href {#1}{\urlprefix }}%
\providecommand \urlprefix  [0]{URL }%
\providecommand \Eprint [0]{\href }%
\providecommand \doibase [0]{http://dx.doi.org/}%
\providecommand \selectlanguage [0]{\@gobble}%
\providecommand \bibinfo  [0]{\@secondoftwo}%
\providecommand \bibfield  [0]{\@secondoftwo}%
\providecommand \translation [1]{[#1]}%
\providecommand \BibitemOpen [0]{}%
\providecommand \bibitemStop [0]{}%
\providecommand \bibitemNoStop [0]{.\EOS\space}%
\providecommand \EOS [0]{\spacefactor3000\relax}%
\providecommand \BibitemShut  [1]{\csname bibitem#1\endcsname}%
\let\auto@bib@innerbib\@empty
\bibitem [{\citenamefont {Hawking}(1975)}]{Hawking1975cmp}%
  \BibitemOpen
  \bibfield  {author} {\bibinfo {author} {\bibfnamefont {S.}~\bibnamefont
  {Hawking}},\ }\href {\doibase 10.1007/BF02345020} {\bibfield  {journal}
  {\bibinfo  {journal} {Commun. Math. Phys.}\ }\textbf {\bibinfo {volume} {43}},\
  \bibinfo {pages} {199} (\bibinfo {year} {1975})}\BibitemShut {NoStop}%
\bibitem [{\citenamefont {Unruh}(1981)}]{Unruh1981prl}%
  \BibitemOpen
  \bibfield  {author} {\bibinfo {author} {\bibfnamefont {W.~G.}\ \bibnamefont
  {Unruh}},\ }\href {\doibase 10.1103/PhysRevLett.46.1351} {\bibfield
  {journal} {\bibinfo  {journal} {Phys. Rev. Lett.}\ }\textbf {\bibinfo {volume}
  {46}},\ \bibinfo {pages} {1351} (\bibinfo {year} {1981})}\BibitemShut
  {NoStop}%
\bibitem [{\citenamefont {Unruh}(1995)}]{Unruh1995prd}%
  \BibitemOpen
  \bibfield  {author} {\bibinfo {author} {\bibfnamefont {W.~G.}\ \bibnamefont
  {Unruh}},\ }\href {\doibase 10.1103/PhysRevD.51.2827} {\bibfield  {journal}
  {\bibinfo  {journal} {Phys. Rev. D}\ }\textbf {\bibinfo {volume} {51}},\
  \bibinfo {pages} {2827} (\bibinfo {year} {1995})}\BibitemShut {NoStop}%
\bibitem [{\citenamefont {Weinfurtner}\ \emph {et~al.}(2011)\citenamefont
  {Weinfurtner}, \citenamefont {Tedford}, \citenamefont {Penrice},
  \citenamefont {Unruh},\ and\ \citenamefont {Lawrence}}]{Weinfurtner2011prl}%
  \BibitemOpen
  \bibfield  {author} {\bibinfo {author} {\bibfnamefont {S.}~\bibnamefont
  {Weinfurtner}}, \bibinfo {author} {\bibfnamefont {E.~W.}\ \bibnamefont
  {Tedford}}, \bibinfo {author} {\bibfnamefont {M.~C.~J.}\ \bibnamefont
  {Penrice}}, \bibinfo {author} {\bibfnamefont {W.~G.}\ \bibnamefont {Unruh}},
  \ and\ \bibinfo {author} {\bibfnamefont {G.~A.}\ \bibnamefont {Lawrence}},\
  }\href {\doibase 10.1103/PhysRevLett.106.021302} {\bibfield  {journal}
  {\bibinfo  {journal} {Phys. Rev. Lett.}\ }\textbf {\bibinfo {volume} {106}},\
  \bibinfo {pages} {021302} (\bibinfo {year} {2011})}\BibitemShut {NoStop}%
\bibitem [{\citenamefont {Eddington}(1920)}]{Eddington1920}%
  \BibitemOpen
  \bibfield  {author} {\bibinfo {author} {\bibfnamefont {A.~S.}\ \bibnamefont
  {Eddington}},\ }\href@noop {} {\emph {\bibinfo {title} {Space, Time, and
  Gravitation}}}\ (\bibinfo  {publisher} {Cambridge University Press, Cambridge, England},\
  \bibinfo {year} {1920})\BibitemShut {NoStop}%
\bibitem [{\citenamefont {Gordon}(1923)}]{Gordon1923}%
  \BibitemOpen
  \bibfield  {author} {\bibinfo {author} {\bibfnamefont {W.}~\bibnamefont
  {Gordon}},\ }\href@noop {} {\bibfield  {journal} {\bibinfo  {journal} {Ann.\
  Phys.}\ }\textbf {\bibinfo {volume} {377}},\ \bibinfo {pages} {421}
  (\bibinfo {year} {1923})}\BibitemShut {NoStop}%
\bibitem [{\citenamefont {Plebanski}(1960)}]{Plebanski1960pr}%
  \BibitemOpen
  \bibfield  {author} {\bibinfo {author} {\bibfnamefont {J.}~\bibnamefont
  {Plebanski}},\ }\href {\doibase 10.1103/PhysRev.118.1396} {\bibfield
  {journal} {\bibinfo  {journal} {Phys. Rev.}\ }\textbf {\bibinfo {volume}
  {118}},\ \bibinfo {pages} {1396} (\bibinfo {year} {1960})}\BibitemShut
  {NoStop}%
\bibitem [{\citenamefont {Felice}(1971)}]{DeFelice1971gerg}%
  \BibitemOpen
  \bibfield  {author} {\bibinfo {author} {\bibfnamefont {F.}~\bibnamefont
  {Felice}},\ }\href {\doibase 10.1007/BF00758153} {\bibfield  {journal}
  {\bibinfo  {journal} {Gen. Relativ. Gravit.}\ }\textbf {\bibinfo {volume} {2}},\
  \bibinfo {pages} {347} (\bibinfo {year} {1971})}\BibitemShut {NoStop}%
\bibitem [{\citenamefont {Rotman}(1962)}]{Rotman1962iretrans}%
  \BibitemOpen
  \bibfield  {author} {\bibinfo {author} {\bibfnamefont {W.}~\bibnamefont
  {Rotman}},\ }\href {\doibase 10.1109/TAP.1962.1137809} {\bibfield  {journal}
  {\bibinfo  {journal} {IEEE Trans. Antennas Propag.}\ }\textbf {\bibinfo
  {volume} {10}},\ \bibinfo {pages} {82 } (\bibinfo {year} {1962})}\BibitemShut
  {NoStop}%
\bibitem [{\citenamefont {Schoenberg}\ and\ \citenamefont
  {Sen}(1983)}]{Schoenberg1983jasa}%
  \BibitemOpen
  \bibfield  {author} {\bibinfo {author} {\bibfnamefont {M.}~\bibnamefont
  {Schoenberg}}\ and\ \bibinfo {author} {\bibfnamefont {P.~N.}\ \bibnamefont
  {Sen}},\ }\href {\doibase 10.1121/1.388724} {\bibfield  {journal} {\bibinfo
  {journal} {J. Acoust. Soc. Am.}\ }\textbf {\bibinfo {volume} {73}},\ \bibinfo
  {pages} {61} (\bibinfo {year} {1983})}\BibitemShut {NoStop}%
\bibitem [{\citenamefont {Pendry}\ \emph {et~al.}(1996)\citenamefont {Pendry},
  \citenamefont {Holden}, \citenamefont {Stewart},\ and\ \citenamefont
  {Youngs}}]{Pendry1996prl}%
  \BibitemOpen
  \bibfield  {author} {\bibinfo {author} {\bibfnamefont {J.~B.}\ \bibnamefont
  {Pendry}}, \bibinfo {author} {\bibfnamefont {A.~J.}\ \bibnamefont {Holden}},
  \bibinfo {author} {\bibfnamefont {W.~J.}\ \bibnamefont {Stewart}}, \ and\
  \bibinfo {author} {\bibfnamefont {I.}~\bibnamefont {Youngs}},\ }\href
  {\doibase 10.1103/PhysRevLett.76.4773} {\bibfield  {journal} {\bibinfo
  {journal} {Phys. Rev. Lett.}\ }\textbf {\bibinfo {volume} {76}},\ \bibinfo
  {pages} {4773} (\bibinfo {year} {1996})}\BibitemShut {NoStop}%
\bibitem [{\citenamefont {Sievenpiper}\ \emph {et~al.}(1996)\citenamefont
  {Sievenpiper}, \citenamefont {Sickmiller},\ and\ \citenamefont
  {Yablonovitch}}]{Sievenpiper1996prl}%
  \BibitemOpen
  \bibfield  {author} {\bibinfo {author} {\bibfnamefont {D.~F.}\ \bibnamefont
  {Sievenpiper}}, \bibinfo {author} {\bibfnamefont {M.~E.}\ \bibnamefont
  {Sickmiller}}, \ and\ \bibinfo {author} {\bibfnamefont {E.}~\bibnamefont
  {Yablonovitch}},\ }\href {\doibase 10.1103/PhysRevLett.76.2480} {\bibfield
  {journal} {\bibinfo  {journal} {Phys. Rev. Lett.}\ }\textbf {\bibinfo
  {volume} {76}},\ \bibinfo {pages} {2480} (\bibinfo {year}
  {1996})}\BibitemShut {NoStop}%
\bibitem [{\citenamefont {Ziolkowski}\ and\ \citenamefont
  {Auzanneau}(1997)}]{Ziolkowski1997jap}%
  \BibitemOpen
  \bibfield  {author} {\bibinfo {author} {\bibfnamefont {R.~W.}\ \bibnamefont
  {Ziolkowski}}\ and\ \bibinfo {author} {\bibfnamefont {F.}~\bibnamefont
  {Auzanneau}},\ }\href {\doibase 10.1063/1.365625} {\bibfield  {journal}
  {\bibinfo  {journal} {J. Appl. Phys.}\ }\textbf {\bibinfo {volume} {82}},\
  \bibinfo {pages} {3195} (\bibinfo {year} {1997})}\BibitemShut {NoStop}%
\bibitem [{\citenamefont {Reznik}(2000)}]{Reznik2000prd}%
  \BibitemOpen
  \bibfield  {author} {\bibinfo {author} {\bibfnamefont {B.}~\bibnamefont
  {Reznik}},\ }\href {\doibase 10.1103/PhysRevD.62.044044} {\bibfield
  {journal} {\bibinfo  {journal} {Phys. Rev. D}\ }\textbf {\bibinfo {volume}
  {62}},\ \bibinfo {pages} {044044} (\bibinfo {year} {2000})}\BibitemShut
  {NoStop}%
\bibitem [{\citenamefont {Sch\"utzhold}\ \emph {et~al.}(2002)\citenamefont
  {Sch\"utzhold}, \citenamefont {Plunien},\ and\ \citenamefont
  {Soff}}]{Schutzhold2002prl}%
  \BibitemOpen
  \bibfield  {author} {\bibinfo {author} {\bibfnamefont {R.}~\bibnamefont
  {Sch\"utzhold}}, \bibinfo {author} {\bibfnamefont {G.}~\bibnamefont
  {Plunien}}, \ and\ \bibinfo {author} {\bibfnamefont {G.}~\bibnamefont
  {Soff}},\ }\href {\doibase 10.1103/PhysRevLett.88.061101} {\bibfield
  {journal} {\bibinfo  {journal} {Phys. Rev. Lett.}\ }\textbf {\bibinfo {volume}
  {88}},\ \bibinfo {pages} {061101} (\bibinfo {year} {2002})}\BibitemShut
  {NoStop}%
\bibitem [{\citenamefont {Greenleaf}\ \emph {et~al.}(2007)\citenamefont
  {Greenleaf}, \citenamefont {Kurylev}, \citenamefont {Lassas},\ and\
  \citenamefont {Uhlmann}}]{Greenleaf2007prl}%
  \BibitemOpen
  \bibfield  {author} {\bibinfo {author} {\bibfnamefont {A.}~\bibnamefont
  {Greenleaf}}, \bibinfo {author} {\bibfnamefont {Y.}~\bibnamefont {Kurylev}},
  \bibinfo {author} {\bibfnamefont {M.}~\bibnamefont {Lassas}}, \ and\ \bibinfo
  {author} {\bibfnamefont {G.}~\bibnamefont {Uhlmann}},\ }\href {\doibase
  10.1103/PhysRevLett.99.183901} {\bibfield  {journal} {\bibinfo  {journal}
  {Phys. Rev. Lett.}\ }\textbf {\bibinfo {volume} {99}},\ \bibinfo
  {pages} {183901} (\bibinfo {year} {2007})}\BibitemShut {NoStop}%
\bibitem [{\citenamefont {Narimanov}\ and\ \citenamefont
  {Kildishev}(2009)}]{Narimanov2009apl}%
  \BibitemOpen
  \bibfield  {author} {\bibinfo {author} {\bibfnamefont {E.~E.}\ \bibnamefont
  {Narimanov}}\ and\ \bibinfo {author} {\bibfnamefont {A.~V.}\ \bibnamefont
  {Kildishev}},\ }\href {\doibase 10.1063/1.3184594} {\bibfield  {journal}
  {\bibinfo  {journal} {Appl Phys Lett}\ }\textbf {\bibinfo {volume} {95}},\
  \bibinfo {eid} {041106} (\bibinfo {year} {2009})}\BibitemShut {NoStop}%
\bibitem [{\citenamefont {Thompson}\ and\ \citenamefont
  {Frauendiener}(2010)}]{Thompson2010prd}%
  \BibitemOpen
  \bibfield  {author} {\bibinfo {author} {\bibfnamefont {R.~T.}\ \bibnamefont
  {Thompson}}\ and\ \bibinfo {author} {\bibfnamefont {J.}~\bibnamefont
  {Frauendiener}},\ }\href {\doibase 10.1103/PhysRevD.82.124021} {\bibfield
  {journal} {\bibinfo  {journal} {Phys.\ Rev.\ D}\ }\textbf {\bibinfo {volume}
  {82}},\ \bibinfo {pages} {124021} (\bibinfo {year} {2010})}\BibitemShut
  {NoStop}%
\bibitem [{\citenamefont {Chen}\ \emph {et~al.}(2010)\citenamefont {Chen},
  \citenamefont {Miao},\ and\ \citenamefont {Li}}]{Chen2010oe}%
  \BibitemOpen
  \bibfield  {author} {\bibinfo {author} {\bibfnamefont {H.}~\bibnamefont
  {Chen}}, \bibinfo {author} {\bibfnamefont {R.-X.}\ \bibnamefont {Miao}}, \
  and\ \bibinfo {author} {\bibfnamefont {M.}~\bibnamefont {Li}},\ }\href
  {\doibase 10.1364/OE.18.015183} {\bibfield  {journal} {\bibinfo  {journal}
  {Opt. Express}\ }\textbf {\bibinfo {volume} {18}},\ \bibinfo {pages} {15183}
  (\bibinfo {year} {2010})}\BibitemShut {NoStop}%
\bibitem [{\citenamefont {Lu}\ \emph {et~al.}(2010)\citenamefont {Lu},
  \citenamefont {Jin}, \citenamefont {Lin},\ and\ \citenamefont
  {Chen}}]{Lu2010jap}%
  \BibitemOpen
  \bibfield  {author} {\bibinfo {author} {\bibfnamefont {W.}~\bibnamefont
  {Lu}}, \bibinfo {author} {\bibfnamefont {J.}~\bibnamefont {Jin}}, \bibinfo
  {author} {\bibfnamefont {Z.}~\bibnamefont {Lin}}, \ and\ \bibinfo {author}
  {\bibfnamefont {H.}~\bibnamefont {Chen}},\ }\href {\doibase
  10.1063/1.3485819} {\bibfield  {journal} {\bibinfo  {journal} {J Appl Phys}\
  }\textbf {\bibinfo {volume} {108}},\ \bibinfo {eid} {064517} (\bibinfo {year}
  {2010})}\BibitemShut {NoStop}%
\bibitem [{\citenamefont {Mackay}\ and\ \citenamefont
  {Lakhtakia}(2010)}]{Mackay2010pla}%
  \BibitemOpen
  \bibfield  {author} {\bibinfo {author} {\bibfnamefont {T.~G.}\ \bibnamefont
  {Mackay}}\ and\ \bibinfo {author} {\bibfnamefont {A.}~\bibnamefont
  {Lakhtakia}},\ }\href {\doibase DOI: 10.1016/j.physleta.2010.03.061}
  {\bibfield  {journal} {\bibinfo  {journal} {Phys Lett A}\ }\textbf {\bibinfo
  {volume} {374}},\ \bibinfo {pages} {2305 } (\bibinfo {year}
  {2010})}\BibitemShut {NoStop}%
\bibitem [{\citenamefont {Miao}\ \emph {et~al.}(2011)\citenamefont {Miao},
  \citenamefont {Zheng},\ and\ \citenamefont {Li}}]{Miao2011plb}%
  \BibitemOpen
  \bibfield  {author} {\bibinfo {author} {\bibfnamefont {R.-X.}\ \bibnamefont
  {Miao}}, \bibinfo {author} {\bibfnamefont {R.}~\bibnamefont {Zheng}}, \ and\
  \bibinfo {author} {\bibfnamefont {M.}~\bibnamefont {Li}},\ }\href {\doibase
  DOI: 10.1016/j.physletb.2011.01.016} {\bibfield  {journal} {\bibinfo
  {journal} {Phys Lett B}\ }\textbf {\bibinfo {volume} {696}},\ \bibinfo
  {pages} {550 } (\bibinfo {year} {2011})}\BibitemShut {NoStop}%
\bibitem [{\citenamefont {Smolyaninov}(2011)}]{Smolyaninov2011jo}%
  \BibitemOpen
  \bibfield  {author} {\bibinfo {author} {\bibfnamefont {I.~I.}\ \bibnamefont
  {Smolyaninov}},\ }\href {\doibase 10.1088/2040-8978/13/2/024004} {\bibfield
  {journal} {\bibinfo  {journal} {J. Opt.}\ }\textbf {\bibinfo
  {volume} {13}},\ \bibinfo {pages} {024004} (\bibinfo {year}
  {2011})}\BibitemShut {NoStop}%
\bibitem [{\citenamefont {Smolyaninov}\ \emph {et~al.}(2012)\citenamefont
  {Smolyaninov}, \citenamefont {Hwang},\ and\ \citenamefont
  {Narimanov}}]{Smolyaninov2012prb}%
  \BibitemOpen
  \bibfield  {author} {\bibinfo {author} {\bibfnamefont {I.~I.}\ \bibnamefont
  {Smolyaninov}}, \bibinfo {author} {\bibfnamefont {E.}~\bibnamefont {Hwang}},
  \ and\ \bibinfo {author} {\bibfnamefont {E.}~\bibnamefont {Narimanov}},\
  }\href {\doibase 10.1103/PhysRevB.85.235122} {\bibfield  {journal} {\bibinfo
  {journal} {Phys. Rev. B}\ }\textbf {\bibinfo {volume} {85}},\ \bibinfo
  {pages} {235122} (\bibinfo {year} {2012})}\BibitemShut {NoStop}%
\bibitem [{\citenamefont {Bini}\ \emph {et~al.}(2014)\citenamefont {Bini},
  \citenamefont {Geralico},\ and\ \citenamefont {Haney}}]{Bini2013gerg}%
  \BibitemOpen
  \bibfield  {author} {\bibinfo {author} {\bibfnamefont {D.}~\bibnamefont
  {Bini}}, \bibinfo {author} {\bibfnamefont {A.}~\bibnamefont {Geralico}}, \
  and\ \bibinfo {author} {\bibfnamefont {M.}~\bibnamefont {Haney}},\ }\href
  {\doibase 10.1007/s10714-013-1644-4} {\bibfield  {journal} {\bibinfo
  {journal} {Gen. Relativ. Gravit.}\ }\textbf {\bibinfo {volume}
  {46}},\ \bibinfo {pages} {1644} (\bibinfo {year} {2014})}\BibitemShut {NoStop}%
\bibitem [{\citenamefont {Boston}(2015)}]{Boston2015prd}%
  \BibitemOpen
  \bibfield  {author} {\bibinfo {author} {\bibfnamefont {S.~R.}\ \bibnamefont
  {Boston}},\ }\href {\doibase 10.1103/PhysRevD.91.124035} {\bibfield
  {journal} {\bibinfo  {journal} {Phys. Rev. D}\ }\textbf {\bibinfo {volume}
  {91}},\ \bibinfo {pages} {124035} (\bibinfo {year} {2015})}\BibitemShut
  {NoStop}%
\bibitem [{\citenamefont {{Fern\'{a}ndez-N\'{u}\~{n}ez}}\ and\ \citenamefont
  {Bulashenko}(2016)}]{FernandezNunez2016pla}%
  \BibitemOpen
  \bibfield  {author} {\bibinfo {author} {\bibfnamefont {I.}~\bibnamefont
  {{Fern\'{a}ndez-N\'{u}\~{n}ez}}}\ and\ \bibinfo {author} {\bibfnamefont
  {O.}~\bibnamefont {Bulashenko}},\ }\href {\doibase
  http://dx.doi.org/10.1016/j.physleta.2015.10.043} {\bibfield  {journal}
  {\bibinfo  {journal} {Phys. Lett. A}\ }\textbf {\bibinfo {volume}
  {380}},\ \bibinfo {pages} {1 } (\bibinfo {year} {2016})}\BibitemShut
  {NoStop}%
\bibitem [{\citenamefont {Pendry}\ \emph {et~al.}(2006)\citenamefont {Pendry},
  \citenamefont {Schurig},\ and\ \citenamefont {Smith}}]{Pendry2006sc}%
  \BibitemOpen
  \bibfield  {author} {\bibinfo {author} {\bibfnamefont {J.~B.}\ \bibnamefont
  {Pendry}}, \bibinfo {author} {\bibfnamefont {D.}~\bibnamefont {Schurig}}, \
  and\ \bibinfo {author} {\bibfnamefont {D.~R.}\ \bibnamefont {Smith}},\ }\href
  {\doibase 10.1126/science.1125907} {\bibfield  {journal} {\bibinfo  {journal}
  {Science}\ }\textbf {\bibinfo {volume} {312}},\ \bibinfo {pages} {1780}
  (\bibinfo {year} {2006})}\BibitemShut {NoStop}%
\bibitem [{\citenamefont {Leonhardt}(2006)}]{Leonhardt2006sc}%
  \BibitemOpen
  \bibfield  {author} {\bibinfo {author} {\bibfnamefont {U.}~\bibnamefont
  {Leonhardt}},\ }\href {\doibase 10.1126/science.1126493} {\bibfield
  {journal} {\bibinfo  {journal} {Science}\ }\textbf {\bibinfo {volume}
  {312}},\ \bibinfo {pages} {1777} (\bibinfo {year} {2006})}\BibitemShut
  {NoStop}%
\bibitem [{\citenamefont {Leonhardt}\ and\ \citenamefont
  {Philbin}(2006)}]{Leonhardt2006njp1}%
  \BibitemOpen
  \bibfield  {author} {\bibinfo {author} {\bibfnamefont {U.}~\bibnamefont
  {Leonhardt}}\ and\ \bibinfo {author} {\bibfnamefont {T.~G.}\ \bibnamefont
  {Philbin}},\ }\href {\doibase 10.1088/1367-2630/8/10/247} {\bibfield
  {journal} {\bibinfo  {journal} {New J. Phys.}\ }\textbf {\bibinfo {volume}
  {8}},\ \bibinfo {pages} {247} (\bibinfo {year} {2006})}\BibitemShut {NoStop}%
\bibitem [{\citenamefont {Rahm}\ \emph {et~al.}(2008)\citenamefont {Rahm},
  \citenamefont {Schurig}, \citenamefont {Roberts}, \citenamefont {Cummer},
  \citenamefont {Smith},\ and\ \citenamefont {Pendry}}]{Rahm2007pn}%
  \BibitemOpen
  \bibfield  {author} {\bibinfo {author} {\bibfnamefont {M.}~\bibnamefont
  {Rahm}}, \bibinfo {author} {\bibfnamefont {D.}~\bibnamefont {Schurig}},
  \bibinfo {author} {\bibfnamefont {D.~A.}\ \bibnamefont {Roberts}}, \bibinfo
  {author} {\bibfnamefont {S.~A.}\ \bibnamefont {Cummer}}, \bibinfo {author}
  {\bibfnamefont {D.~R.}\ \bibnamefont {Smith}}, \ and\ \bibinfo {author}
  {\bibfnamefont {J.~B.}\ \bibnamefont {Pendry}},\ }\href {\doibase DOI:
  10.1016/j.photonics.2007.07.013} {\bibfield  {journal} {\bibinfo  {journal}
  {Phot. Nano. Fund. Appl.}\ }\textbf {\bibinfo {volume} {6}},\ \bibinfo
  {pages} {87 } (\bibinfo {year} {2008})}\BibitemShut {NoStop}%
\bibitem [{\citenamefont {Post}(1962)}]{Post}%
  \BibitemOpen
  \bibfield  {author} {\bibinfo {author} {\bibfnamefont {E.~J.}\ \bibnamefont
  {Post}},\ }\href@noop {} {\emph {\bibinfo {title} {Formal Structure of
  Electromagnetics}}}\ (\bibinfo  {publisher} {North-Holland},\
  \bibinfo {address} {Amsterdam},\ \bibinfo {year} {1962})\BibitemShut
  {NoStop}%
\bibitem [{\citenamefont {Misner}\ \emph {et~al.}(1973)\citenamefont {Misner},
  \citenamefont {Thorne},\ and\ \citenamefont {Wheeler}}]{MTW}%
  \BibitemOpen
  \bibfield  {author} {\bibinfo {author} {\bibfnamefont {C.~W.}\ \bibnamefont
  {Misner}}, \bibinfo {author} {\bibfnamefont {K.~S.}\ \bibnamefont {Thorne}},
  \ and\ \bibinfo {author} {\bibfnamefont {J.~A.}\ \bibnamefont {Wheeler}},\
  }\href@noop {} {\emph {\bibinfo {title} {Gravitation}}}\ (\bibinfo
  {publisher} {Freeman},\ \bibinfo {address} {San Francisco},\
  \bibinfo {year} {1973})\BibitemShut {NoStop}%
\bibitem [{\citenamefont {Baez}\ and\ \citenamefont {Muniain}(1994)}]{Baez}%
  \BibitemOpen
  \bibfield  {author} {\bibinfo {author} {\bibfnamefont {J.}~\bibnamefont
  {Baez}}\ and\ \bibinfo {author} {\bibfnamefont {J.~P.}\ \bibnamefont
  {Muniain}},\ }\href@noop {} {\emph {\bibinfo {title} {Gauge Fields, Knots and
  Gravity}}}\ (\bibinfo  {publisher} {World Scientific},\ \bibinfo {address}
  {Singapore},\ \bibinfo {year} {1994})\BibitemShut {NoStop}%
\bibitem [{\citenamefont {Thompson}\ and\ \citenamefont
  {Cummer}(2012)}]{Thompson2012aiep}%
  \BibitemOpen
  \bibfield  {author} {\bibinfo {author} {\bibfnamefont {R.~T.}\ \bibnamefont
  {Thompson}}\ and\ \bibinfo {author} {\bibfnamefont {S.~A.}\ \bibnamefont
  {Cummer}},\ }in\ \href {\doibase 10.1016/B978-0-12-394297-5.00005-2} {\emph
  {\bibinfo {booktitle} {Advances in Imaging and Electron Physics}}},\ Vol.\
  \bibinfo {volume} {171},\ \bibinfo {editor} {edited by\ \bibinfo {editor}
  {\bibfnamefont {P.~W.}\ \bibnamefont {Hawkes}}}\ (\bibinfo  {publisher}
  {Elsevier, New York},\ \bibinfo {year} {2012}),\ p.\ \bibinfo {pages}
  {195}\BibitemShut {NoStop}%
\bibitem [{\citenamefont {Thompson}\ \emph
  {et~al.}(2011{\natexlab{a}})\citenamefont {Thompson}, \citenamefont
  {Cummer},\ and\ \citenamefont {Frauendiener}}]{Thompson2011jo1}%
  \BibitemOpen
  \bibfield  {author} {\bibinfo {author} {\bibfnamefont {R.~T.}\ \bibnamefont
  {Thompson}}, \bibinfo {author} {\bibfnamefont {S.~A.}\ \bibnamefont
  {Cummer}}, \ and\ \bibinfo {author} {\bibfnamefont {J.}~\bibnamefont
  {Frauendiener}},\ }\href {\doibase 10.1088/2040-8978/13/2/024008} {\bibfield
  {journal} {\bibinfo  {journal} {J.\ Opt.}\ }\textbf {\bibinfo {volume}
  {13}},\ \bibinfo {pages} {024008} (\bibinfo {year}
  {2011}{\natexlab{a}})}\BibitemShut {NoStop}%
\bibitem [{\citenamefont {Thompson}\ \emph
  {et~al.}(2011{\natexlab{b}})\citenamefont {Thompson}, \citenamefont
  {Cummer},\ and\ \citenamefont {Frauendiener}}]{Thompson2011jo2}%
  \BibitemOpen
  \bibfield  {author} {\bibinfo {author} {\bibfnamefont {R.~T.}\ \bibnamefont
  {Thompson}}, \bibinfo {author} {\bibfnamefont {S.~A.}\ \bibnamefont
  {Cummer}}, \ and\ \bibinfo {author} {\bibfnamefont {J.}~\bibnamefont
  {Frauendiener}},\ }\href {\doibase 10.1088/2040-8978/13/5/055105} {\bibfield
  {journal} {\bibinfo  {journal} {J.\ Opt}\ }\textbf {\bibinfo {volume} {13}},\
  \bibinfo {pages} {055105} (\bibinfo {year} {2011}{\natexlab{b}})}\BibitemShut
  {NoStop}%
\bibitem [{\citenamefont {Ryder}(2009)}]{Ryder2009}%
  \BibitemOpen
  \bibfield  {author} {\bibinfo {author} {\bibfnamefont {L.}~\bibnamefont
  {Ryder}},\ }\href {http://dx.doi.org/10.1017/CBO9780511809033} {\emph
  {\bibinfo {title} {Introduction to General Relativity}}}\ (\bibinfo
  {publisher} {Cambridge University Press, Cambridge, England},\ \bibinfo {year} {2009}).
\bibitem [{\citenamefont {Cummer}\ and\ \citenamefont
  {Thompson}(2011)}]{Cummer2011jo}%
  \BibitemOpen
  \bibfield  {author} {\bibinfo {author} {\bibfnamefont {S.~A.}\ \bibnamefont
  {Cummer}}\ and\ \bibinfo {author} {\bibfnamefont {R.~T.}\ \bibnamefont
  {Thompson}},\ }\href {\doibase 10.1088/2040-8978/13/2/024007} {\bibfield
  {journal} {\bibinfo  {journal} {J. Opt.}\ }\textbf {\bibinfo {volume} {13}},\
  \bibinfo {pages} {024007} (\bibinfo {year} {2011})}\BibitemShut {NoStop}%
\bibitem [{\citenamefont {Poisson}(2004)}]{Poisson}%
  \BibitemOpen
  \bibfield  {author} {\bibinfo {author} {\bibfnamefont {E.}~\bibnamefont
  {Poisson}},\ }\href {http://dx.doi.org/10.1017/CBO9780511606601} {\emph
  {\bibinfo {title} {A Relativist's Toolkit}}}\ (\bibinfo  {publisher}
  {Cambridge University Press, Cambridge, England},\ \bibinfo {year} {2004}).
\bibitem [{\citenamefont {Faroni}(2015)}]{Faroni2015}%
  \BibitemOpen
  \bibfield  {author} {\bibinfo {author} {\bibfnamefont {V.}~\bibnamefont
  {Faroni}},\ }\href {http://dx.doi.org/10.1007/978-3-319-19240-6} {\emph
  {\bibinfo {title} {Cosmological and Black Hole Apparent Horizons}}} (\bibinfo  {publisher} {Springer,
  New York},\ \bibinfo {year} {2015})\BibitemShut {NoStop}%
\bibitem [{\citenamefont {Thompson}\ and\ \citenamefont
  {Fathi}(2016)}]{Thompson2016}%
  \BibitemOpen
  \bibfield  {author} {\bibinfo {author} {\bibfnamefont {R.~T.}\ \bibnamefont
  {Thompson}}\ and\ \bibinfo {author} {\bibfnamefont {M.}~\bibnamefont
  {Fathi}}\ }\href@noop {} {\bibfield  {journal} {\bibinfo  {journal} {(to be published).}\ }}
\bibitem [{\citenamefont {Thompson}\ and\ \citenamefont
  {Fathi}(2015)}]{Thompson2015pra}%
  \BibitemOpen
  \bibfield  {author} {\bibinfo {author} {\bibfnamefont {R.~T.}\ \bibnamefont
  {Thompson}}\ and\ \bibinfo {author} {\bibfnamefont {M.}~\bibnamefont
  {Fathi}},\ }\href {\doibase 10.1103/PhysRevA.92.013834} {\bibfield  {journal}
  {\bibinfo  {journal} {Phys.\ Rev.\ A}\ }\textbf {\bibinfo {volume} {92}},\
  \bibinfo {pages} {013834} (\bibinfo {year} {2015})}\BibitemShut {NoStop}%
\bibitem [{\citenamefont {Thompson}(2010)}]{Thompson2010pra}%
  \BibitemOpen
  \bibfield  {author} {\bibinfo {author} {\bibfnamefont {R.~T.}\ \bibnamefont
  {Thompson}},\ }\href {\doibase 10.1103/PhysRevA.82.053801} {\bibfield
  {journal} {\bibinfo  {journal} {Phys.\ Rev.\ A}\ }\textbf {\bibinfo {volume}
  {82}},\ \bibinfo {pages} {053801} (\bibinfo {year} {2010})}\BibitemShut
  {NoStop}%
\bibitem [{\citenamefont {Thompson}(2012)}]{Thompson2012jo1}%
  \BibitemOpen
  \bibfield  {author} {\bibinfo {author} {\bibfnamefont {R.~T.}\ \bibnamefont
  {Thompson}},\ }\href {\doibase 10.1088/2040-8978/14/1/015102} {\bibfield
  {journal} {\bibinfo  {journal} {J.\ Opt.}\ }\textbf {\bibinfo {volume}
  {14}},\ \bibinfo {pages} {015102} (\bibinfo {year} {2012})}\BibitemShut
  {NoStop}%
\end{thebibliography}

%

\end{document}